\documentclass[floatfix,prl,twocolumn,aps,superscriptaddress,showpacs]{revtex4-1}

\usepackage[colorlinks=true,linktoc=page,linkcolor=PineGreen,citecolor=magenta,urlcolor=violet]{hyperref}
\usepackage{multirow}
\usepackage{adjustbox}
\usepackage{graphicx}
\usepackage{amsmath}
\usepackage{physics}
\usepackage{amsfonts}
\usepackage{amssymb}
\usepackage{bm}
\usepackage[usenames,dvipsnames]{color}
\usepackage{float}
\usepackage{natbib}
\usepackage{comment}
\hypersetup{
colorlinks=true,
citecolor=blue,
linkcolor=blue,
urlcolor=blue}

\newcommand{\be}{\begin{equation}}
\newcommand{\ee}{\end{equation}}
\newcommand{\bea}{\begin{eqnarray}}
\newcommand{\eea}{\end{eqnarray}}
\newcommand{\bec}{\begin{center}}
\newcommand{\eec}{\end{center}}

\usepackage[dvipsnames]{xcolor}
\usepackage{multirow}
\usepackage{xfrac}
\usepackage{textcomp}
\usepackage{enumerate}
\usepackage{subfigure}
\usepackage{stackengine}
\usepackage{booktabs}
\usepackage{physics}
\usepackage[colorlinks=true,linktoc=page,linkcolor=PineGreen,citecolor=magenta,urlcolor=violet]{hyperref}
\usepackage{newfloat}
\DeclareFloatingEnvironment[name={TABLE},fileext=lsst,listname={List of
	Supplementary Short Tables},within=section, placement=htbp!]{supptable}

\mathchardef\mhyphen="2D 

\newcommand\beq{\begin{equation}}  
\newcommand\eeq{\end{equation}}

\usepackage[normalem]{ulem}
\definecolor{lime}{HTML}{A6CE39}
\usepackage{sidecap,tikz}
\DeclareRobustCommand{\orcidicon}{\hspace{-1.0mm}
	\begin{tikzpicture}
	\draw[lime, fill=lime] (0.0,0.0) 
	circle [radius=0.15] 
	node[white] {{\fontfamily{qag}\selectfont \tiny \,ID}};
	\draw[white, fill=white] (-0.0525,0.095) 
	circle [radius=0.007];
	\end{tikzpicture}
	\hspace{-3.0mm}
}
\foreach \x in {A, ..., Z}{\expandafter\xdef\csname orcid\x\endcsname{\noexpand\href{https://orcid.org/\csname orcidauthor\x\endcsname}
		{\noexpand\orcidicon}}
}

\begin{document}

\title{
Skyrmion-Antiskyrmion Lattice: A \textit{Net-Zero} Topological Phase in Low-Symmetry Frustrated Chiral Magnets 
}

\author{Sayan Banik\orcidA{}}
\affiliation{School of Physical Sciences, National Institute of Science Education and Research, An OCC of Homi Bhabha National Institute, Jatni-752050, India}
\author{Ashis K.\ Nandy\orcidB{}}
\email{aknandy@niser.ac.in}
\affiliation{School of Physical Sciences, National Institute of Science Education and Research, An OCC of Homi Bhabha National Institute, Jatni-752050, India}

\begin{abstract}
We report the discovery of a thermodynamically stable skyrmion-antiskyrmion lattice in two-dimensional heterostructures, a novel state exhibiting a \textit{net-zero} global topological charge owing to an equal population of skyrmions and antiskyrmions. 
This surprising coexistence of oppositely charged solitons remarkably circumvents their anticipated annihilation.
We demonstrate the formation and evolution of this phase in Fe films on $C_{1v}$-symmetric (110) surfaces of GaAs and CdTe semiconductors. Specifically, we reveal a series of magnetic field-induced phase transitions: cycloidal spin-spiral $\rightarrow$ skyrmyion-antiskyrmion lattice $\rightarrow$ conical spin-spiral $\rightarrow$ ferromagnet.
The remarkable stability of the \textit{net-zero} lattice is attributed to symmetry-enforced anisotropic magnetic interactions. 
Lowering interfacial symmetry to $C_{1v}$ thus enables frustrated chiral magnets, uniquely manifesting in thermodynamically stable \textit{net-zero} topological soliton lattices, as revealed by our findings. 

\end{abstract}

\maketitle

\noindent\textcolor{blue}{\textit{Introduction.-}} 
Magnetic solitons characterized by a topological charge (TC) $Q${--}often referred to as particle-like skyrmions~\cite{Bogdanov_89, Bogdanov_99, Bogdanov_1994JMMM, Bogdanov_1994, Heo_SciRep2016, Dupe-NatCommun2016, Flovik_PRBR2017, Rybakov_19, Foster_19, Nandy_NanoLett2020,  Yang_24embedded, Kuchkin_20i, Kuchkin_20ii, Kuchkin_21, Kuchkin_22homotopy, Kuchkin_23}{--}arise from competing energies: Heisenberg exchange, Dzyaloshinskii-Moriya (DM)~\cite{DMI1,DMI2} interactions, magnetocrystalline anisotropy, and the Zeeman effect. 
Their topological protection ensures stability against deformation into trivial states, such as uniform magnetization. 
Isolated magnetic skyrmions with $Q=-1$ TC can be stabilized as both static and dynamic magnetic quasiparticles through local energy minima~\cite{Bogdanov_1994, Heo_SciRep2016, Flovik_PRBR2017, Dupe-NatCommun2016, Romming_PRL15, Nandy_NanoLett2020}.
Nevertheless, an external magnetic field can drive a phase transition from the chiral spin-spiral (SS) to a stable equilibrium phase, an ordered skyrmion lattice (SkL), as a global energy minimum.
The subsequent discovery of antiskyrmion lattices \cite{Nayak-Nature2017}, composed of antiskyrmions ($Q = +1$) with TC opposite to skyrmions, has broadened the variety of homogeneous magnetic soliton lattices.
Extensive studies across bulk noncentrosymmetric systems underscore the crucial role of crystal symmetry in determining the field-induced equilibrium lattice phases{--}specifically, either SkL ($T$, $O$ and $C_{nv}$ point groups)~\cite{Muehlbauer-Science2009, Yu_10, Yu_11, Seki_12multiferroic, Kezsmarki_15polar} or antiskyrmion lattice ($D_{2d}$ and S$_4$ point groups) ~\cite{Nayak-Nature2017, Jena-NatCommun_2020, Karube-NatMat_2021}. 
In two-dimensional (2D) settings, chiral magnets mostly realized in interfacial transition-metal/heavy-metal (TM/HM) heterostructures offer a highly tunable platform for stabilizing and manipulating nanoscale magnetic skyrmions~\cite{Heinze_NatPhys2011, Romming_13, Dupe_NatCommun2014, FerrianiPRL, Nandy_PRL2016}.
Distinct from bulk chiral magnets, these systems feature a unique synergy between exchange frustration and interfacial DM interaction (iDMI), stabilizing frustrated chiral SSs, such as the cycloidal spin spiral (CySS) phase.
Notable examples include square ($C_{4v}$; Fig.~\ref{fig_1}(a))~\cite{FerrianiPRL, Nandy_PRL2016, Nandy-PRBL2024} and triangular ($C_{3v}$; Fig.~\ref{fig_1}(b))~\cite{Heinze_NatPhys2011, Romming_13, Dupe_NatCommun2014, Romming_PRL15, Dupe-NatCommun2016} interfacial geometries, both of which exhibit a characteristic field-driven phase transition sequence: from a CySS to SkL, and ultimately to a saturated ferromagnet (FM).
Recent predictions reveal metastable antiskyrmions in 2Fe/W(110)~\cite{Hoffmann-NatCommun2017} under rectangular symmetry ($C_{2v}$; Fig.~\ref{fig_1}(c)), stabilized by $D_{2d}$-symmetric DM vectors. Although this symmetry is typical in bulk chiral magnets, here it effectively results from the addition of iDMI within the layers.

While the symmetry of DM vectors typically dictates the stabilization of equilibrium lattices composed of either skyrmions or antiskyrmions, a fundamental question remains:  Can a coexisting skyrmion-antiskyrmion lattice (Sk-ASkL) be realized with balanced populations of $Q= \pm 1$ quasiparticles? 
If so, the exact cancellation of TC would result in a magnetic phase with vanishing global topology{--}a state defined as \textit{net-zero} TC lattice with unit-cell charge $Q_\textrm{UC}$=0.
As reported recently by Pham et al.~\cite{Pham_Science2024}, vertically stacked, antiferromagnetically (AFM) coupled magnetic layers ($e.g.$, synthetic antiferromagnets) can host isolated \textit{net-zero} quasiparticles formed by skyrmion pairs. 
A key feature of this quasiparticle is the suppression of the skyrmion Hall effect, resulting from the topological charge cancellation between pairs of opposite polarity.
Notably, the dynamic creation and annihilation of $Q=\pm 1$ pairs, a phenomenon akin to particle-antiparticle pair behavior, has been reported  previously~\cite{Ritzmann_NatElctron2018}.
Therefore, even considering the remarkable advantage of \textit{net-zero} TC spin textures, achieving a thermodynamically stable Sk-ASkL in single-layer magnets remains counterintuitive, as the pair annihilation challenges the stabilization of their coexistence~\cite{Zheng_22, Kuchkin_20i}.

\begin{figure*}[ht]
    \centering \includegraphics[width=0.9\linewidth]{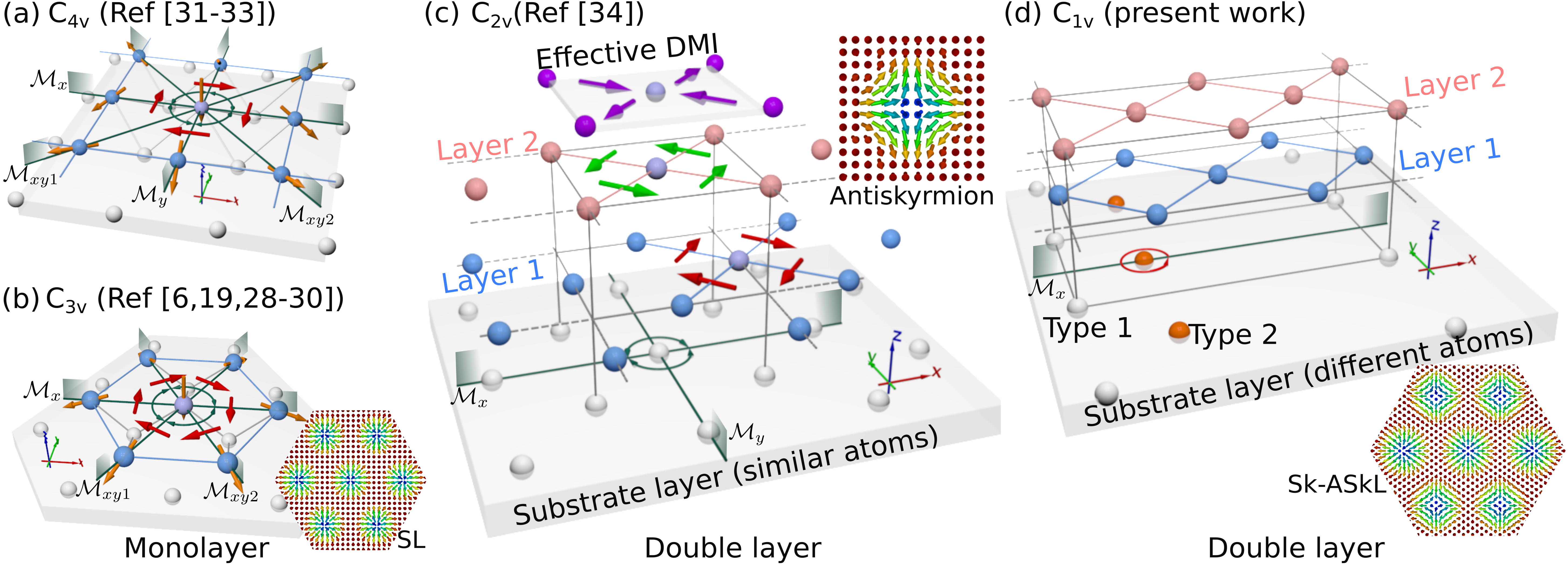}
    \caption{Symmetry-dependent N\'eel-type topological spin configurations in interfacial chiral magnets. Schematic illustrations depict chiral magnetic heterostructures with interfacial symmetries: (a) four-fold ($C_{4v}$), (b) three-fold ($C_{3v}$), and (c) two-fold ($C_{2v}$) and (d) single-fold ($C_{1v}$, the present work). Mirror planes ($\mathcal{M}$) are indicated. For the well-known $C_{4v}$ and $C_{3v}$ systems, symmetry-restricted DM vectors (Moriya rules) lead to stabilize hexagonal SkLs. The $C_{2v}$ interface, exhibiting effective $D_{2d}$ symmetry, stabilizes isolated antiskyrmions. Unlike typical TM/HM heterostructures, we investigate magnetic thin-film interfaced with semiconductor substrates, possessing $C_{1v}$ symmetry. The corresponding magnetic model is described in Eq.~\ref{spin_hamiltonian}. 
    }
    \label{fig_1}
\end{figure*}

In this Letter, we predict that the Sk-ASkL emerges as a thermodynamically stable phase induced by a magnetic field in $C_{1v}$-symmetric transition-metal/semiconductor (TM/SC) heterostructures. 
Archetypal systems for these phases include interfacial magnetic films{--}transition metals ($e.g.$, Fe) on zincblende (110) semiconductor substrates ($e.g.$, GaAs and CdTe).
A chiral CySS with nanometer-scale period ($\lambda)$ forms the zero-field ground state, stabilized by the interplay of interactions, including exchange frustration and iDMI, and out-of-plane uniaxial anisotropy.
An external magnetic field induces a sequence of phase transitions: from CySS to the Sk-ASkL, then to a conical spin-spiral (CoSS) state, and ultimately to a saturated FM state. Remarkably, the anisotropic exchange parameter space in chiral magnets, resulting from the symmetry reduction to $C_{1v}$ (Fig.~\ref{fig_1}(d)), predicts the formation of the 
\textit{net-zero} TC lattice. 
Notably, this equilibrium phase is fundamentally distinct from systems where \textit{net-zero} topology arises from same-type quasiparticles in bilayer geometries~\cite{Zhang_NatCommun2016, Zhang_NanoLett2024}. Here, the Sk-ASkL is characterized by the spontaneous nucleation of skyrmions and antiskyrmions within a single ferromagnetic layer, stabilized purely by the intrinsic balance of competing magnetic interactions.

\begin{figure}[ht]
    \centering \includegraphics[width=0.95\linewidth]{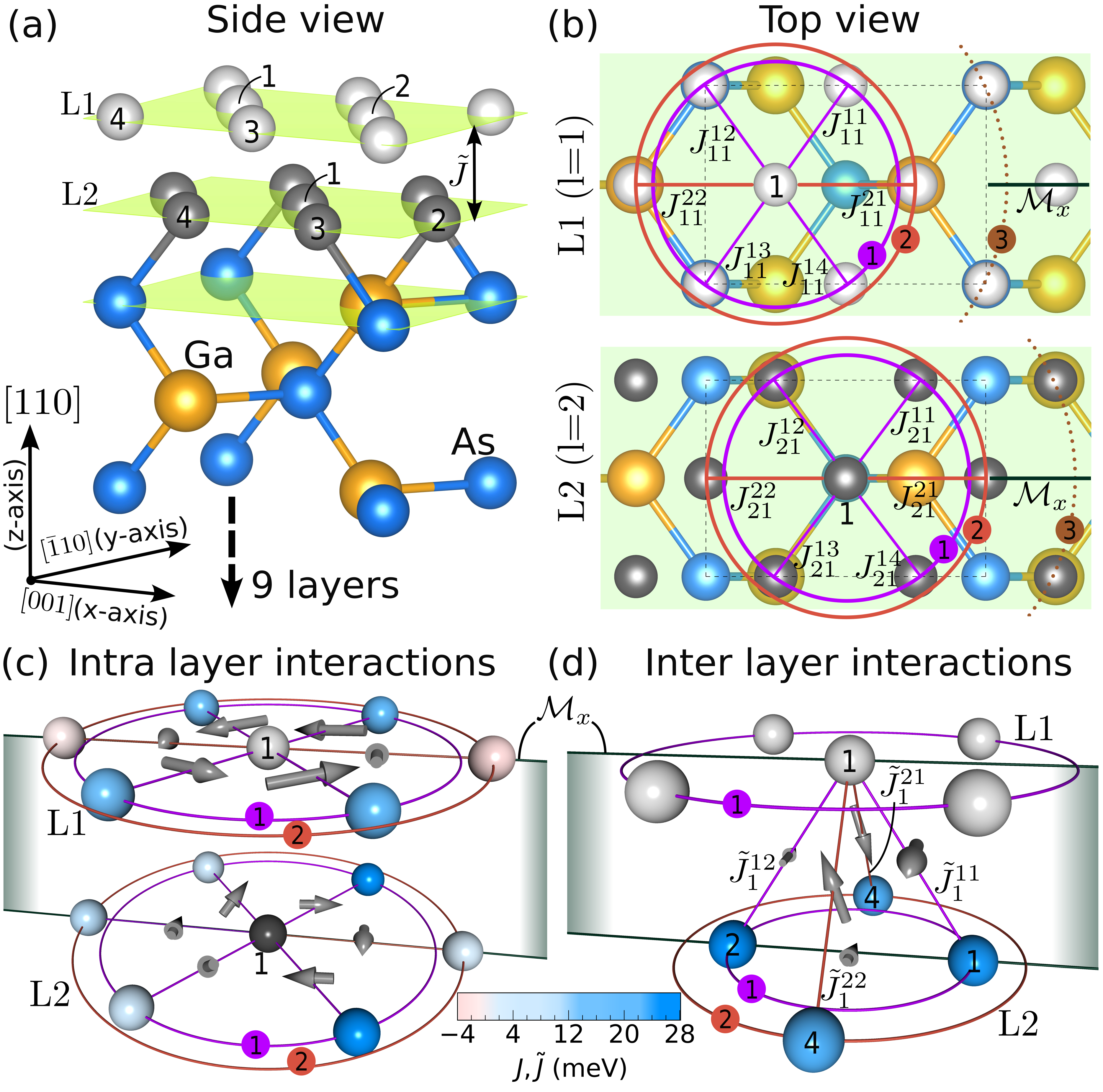}
    \caption{A typical $C_{1v}$ symmetric interface geometry: 2Fe/GaAs(110) slab (a) Side view of the epitaxial Fe double-layer slab, highlighting inequivalent Fe atoms (indexed $\alpha = 1-4$) in surface (L1) and interface (L2) layers.  
    (b) Top-view of L1 and L2, emphasizing asymmetric coordination environments induced by the substrate's low symmetry. A single mirror plane, $\mathcal{M}_x$ (aligned with [001], $x$-axis), governs symmetry constraints. First three neighboring shells illustrate spatially varying intralayer exchange strengths (FM/AFM) due to inequivalent Fe-substrate bonding. (c) Intralayer and (d) interlayer interaction shells near the central Fe atom  ($\alpha=1$), mapping exchange constants (color scale: FM = $+$ve, AFM = $-$ve) and DM vectors (arrows). DM interactions within the $\mathcal{M}_x$ plane exhibit purely $y$-axis components (see also SM Secs. III (2Fe/GaAs(110)) and IV (2Fe/CdTe(110))~\cite{supp})
}
    \label{fig_2}
\end{figure}

\noindent \textcolor{blue}{\textit{Magnetic heterostructure design.-}}  
Utilizing the (110) surface of binary SCs as a cleavage plane enables the formation of magnetic heterostructures where the interfacial symmetry is reduced to $C_{1v}$ symmetry (see Fig.~\ref{fig_1} for comparison).
The 2Fe/GaAs(110) heterostructure, a two-monolayer-thick Fe film with a rectangular unit cell grown epitaxially on GaAs(110), is depicted in Fig.~\ref{fig_2}(a).
Atomic positions within the surface (L1) and interface (L2) Fe layers are explicitly labeled. The GaAs lattice constant is held at its experimental value of 5.656 \AA, defining the \textit{pristine} slab geometry. 
The inherent $C_{1v}$ symmetry of the (110) substrate is preserved, featuring a mirror $\mathcal{M}_x$ parallel to the $x$-axis ($i.e.$, the [001] crystal axis).
The nearly doubled lattice constant of GaAs compared to $\alpha$-Fe results in a close lattice match ($\sim -1.4\%$, compressive strain) between the substrate and the Fe layer, enabling high-quality epitaxial growth~\cite{GaAs(110)_PRB, Carbone_SSC87, Ifflander_PRL2015}.
However, the difference in planar atom density, where two inequivalent substrate atoms (Ga and As) occupy an area equivalent to four Fe atoms, leads to varied chemical environments, see Fig.~\ref{fig_2}(b).

The CdTe SC substrate, a II-VI example system with a lattice parameter 6.478~\AA (exceeding that of GaAs), introduces a substantial lattice mismatch ($\sim 11.4\%$, tensile strain) at the interface. Note, similar tensile strain has been found to persist in Fe/InAs(110) sample~\cite{Sacharow_PRB04}).
Following atomic position relaxation, structural and morphological analyses reveal interfacial roughness, consistent with inherent (110) SC surface corrugation~\cite{Feenstra_PRB1985,Pessa_jap1983}.
Thus, in contrast to the smooth interfaces of TM/HM chiral magnets, multi-atom-per-layer slabs in our systems introduce significant interfacial uniqueness and complexity.
The relaxed interface geometry of our slabs is detailed in the Supplementary Material (SM) Sec. II~\cite{supp}.

\noindent \textcolor{blue}{\textit{Atomistic multilayer model.-}} 
Consistent with earlier atomically thin TM/HM models, the Heisenberg Hamiltonian employed for our TM/SC magnetic heterostructures is given by:
\begin{align}
\label{spin_hamiltonian}
    \mathcal{H} &=\mathcal{H}^{\textrm{Intra}} + \mathcal{H}^{\textrm{Inter}} + \mathcal{H}^{\textrm{A}} + \mathcal{H}^{\textrm{Z}}\\  \nonumber
   &=-\frac{1}{2}\sum_{sl\alpha\beta}\! \bigg[\!{J}^{s\beta}_{l\alpha}~\hat{\bm{n}}_{l\alpha} \cdot \hat{\bm{n}}_{l\beta} + \bm{D}^{s\beta}_{l\alpha} \cdot \Big(\hat{\bm{n}}_{l\alpha} \times \hat{\bm{n}}_{l\beta}\Big)\bigg] \\ \nonumber
   &~~~\!-\sum_{s\alpha\beta}\!\bigg[\!\tilde{J}^{s\beta}_{\alpha} \hat{\bm{n}}_{1\alpha} \cdot \hat{\bm{n}}_{2\beta} +  \bm{\tilde{D}}^{s\beta}_{\alpha} \cdot \Big(\hat{\bm{n}}_{1\alpha} \times \hat{\bm{n}}_{2\beta}\Big)\bigg] \\ \nonumber
   &~~~\!-\textrm{K} \sum_{i}(\hat{\bm{n}}_i \cdot \hat{z})^2 -\mu_s\sum_{i} \textbf{B}_\textrm{ext} \cdot {\bm{\hat{n}}_i}~~.
\end{align} 
The first two terms, $\mathcal{H}^\textrm{Intra}$ and $\mathcal{H}^\textrm{Inter}$, describe chiral Hamiltonians, incorporating intra- and interlayer exchange couplings ($J$, $\tilde{J}$) and DM vectors ($\mathbf{D}$, $\mathbf{\tilde{D}}$). 
The third term, $\mathcal{H}^\textrm{A}$, quantifies the uniaxial anisotropy energy (K), while the last term, $\mathcal{H}^\textrm{Z}$, corresponds to the Zeeman energy induced by an external magnetic field (\textbf{B}$_\textrm{ext}$). 
The exchange and DM vectors are made site-specific through indexing, thereby reflecting the influence of chemical inhomogeneity on the magnetic interactions.
The interaction parameters are explicitly indexed by $\alpha$ and $\beta$, which correspond to the atomic positions within the unit cell of the $l$-th layer (L1 or L2) and the $s$-th neighboring shell of each Fe atom, respectively. 
The magnetization direction is given by $\hat{\bm{n}}$. We derive all material parameters within density functional theory utilizing the Korringa-Kohn-Rostoker (KKR) Green function approach~\cite{KKR_Code}. Computational details are provided in the SM, Sec. I~\cite{supp}.

\noindent \textcolor{blue}{\textit{Material specific model parameters.-}} The computed parameter space is found to be distinct from that of the well-studied high-symmetry configurations, $C_{4v}$ and $C_{3v}$. 
Specifically, triangular and square lattices display isotropic interaction profiles: exchange couplings within a neighboring shell are uniformly FM or AFM, and DM vectors are strictly confined to the lattice plane, adhering to Moriya's rules (see Figs.~\ref{fig_1}(a) and (b)).
The $C_{1v}$ symmetry, lacking the $\mathcal{M}_y$ mirror plane, uniquely manifests a left-right asymmetry. 
This asymmetry leads to a marked anisotropy in magnetic interactions, wherein the exchange and DM vectors are directionally dependent.

For instance, as illustrated in Fig.~\ref{fig_2}(b), the atom designated $\alpha=1$ exhibits intralayer exchange parameters associated with its first and second nearest-neighbor atoms.
As shown in Fig.~\ref{fig_2}(c), the sphere color saturation indicates inequivalent exchange parameters within a given shell: $J_{11}^{11} (=J_{11}^{14}) \neq J_{11}^{12} (=J_{11}^{13})$ for shell $s=1$, and $J_{11}^{21} \neq J_{11}^{22}$ for shell $s=2$. 
Blue and red spheres denote FM and AFM couplings, respectively, where saturation intensity scales with interaction strength.
As depicted in Fig.~\ref{fig_2}(d), the interlayer FM exchange couplings are found to be dominant. 
Quantitatively, for 2Fe/GaAs(110), the largest intralayer and interlayer FM couplings are 28.1 meV and 37.7 meV, respectively. 
For 2Fe/CdTe(110), the interlayer coupling (50.4 meV) is more than double the intralayer coupling (23.6 meV).
Detailed Fe-Fe exchange interaction analysis indicates pronounced magnetic frustration, with dominant FM couplings in the first few neighbor shells (see Figs.~S2 (2Fe/GaAs(110)) and S5 (2Fe/CdTe(110)) in the SM~\cite{supp}).

Visual inspection of the intralayer and interlayer DM vector orientations, depicted as arrows in Figs.~\ref{fig_2}(c) and (d), clearly indicates the dominant contribution from the first two coordination shells.
The strength of the DM interaction is observed to diminish rapidly with distance.
For intralayer nearest-neighbor interactions ($s=1)$, the symmetry reduction results in a non-orthogonal alignment of the DM vectors, $\mathbf{D}^{1\beta}_{11}$ ($\beta=1{-}4$), relative to the bond directions. They also feature a small out-of-plane component.
In contrast, both the intralayer next-nearest-neighbor ($s=2$) and interlayer nearest-neighbor ($s=1$) DM vectors remain strictly in-plane, exhibiting only a $y$-component. This in-plane orientation and specific component are consistent with Moriya's symmetry rules~\cite{DMI2}, attributed to the connecting atoms lying within the $\mathcal{M}_x$ mirror plane.
The remaining interlayer DM vectors in $s=2$ are primarily out-of-plane, with moderate in-plane contributions.
The DM vector orientations for other nonequivalent Fe atoms are presented in the insets of Figs.~S2 and S5.
Crucially, the direction-dependent anisotropic nature of these DM vectors directly stems from the the asymmetric substrate atomic environment around each magnetic atom.
The symmetry breaking inherent in the $C_{1v}$ systems also results in a notable directional anisotropy of the DM vector magnitudes, $e.g.$, for inplane vectors, $|\mathbf{D}^{21}_{11}|\ne |\mathbf{D}^{22}_{11}|$, $|\mathbf{D}^{21}_{21}|\ne |\mathbf{D}^{22}_{21}|$ and $|\tilde{\mathbf{D}}^{11}_{1}|\ne |\tilde{\mathbf{D}}^{12}_{1}|$.
As an example, if an Fe atom in Layer L1 (indexed `1' in Figure 2(b) and assumed at the origin) has an adjacent As substrate atom at $+x$, the lack of a corresponding atom at $-x$ directly causes the inequality $D_{11}^{21} \ne D_{11}^{22}$.
This anisotropy, present in all DM vectors within a shell, contrasts with the isotropy found in $C_{3v}$ and $C_{4v}$ systems and distinguishes TM/SC (110) interfaces from TM/HM interfaces.
Furthermore, the opposing orientation of DM vectors across layers leads to partial cancellation, effectively suppressing the resultant iDMI strength in the $\mathcal{H}^{\textrm{Intra}}$ term. 
While we present the 2Fe/GaAs(110) case here, analogous magnetic interaction landscapes are observed for the 2Fe/CdTe(110) system (as detailed in SM, Sec. IV~\cite{supp}), and both systems exhibit out-of-plane magnetocrystalline anisotropy.

\begin{figure}
    \centering    \includegraphics[width=0.95\linewidth]{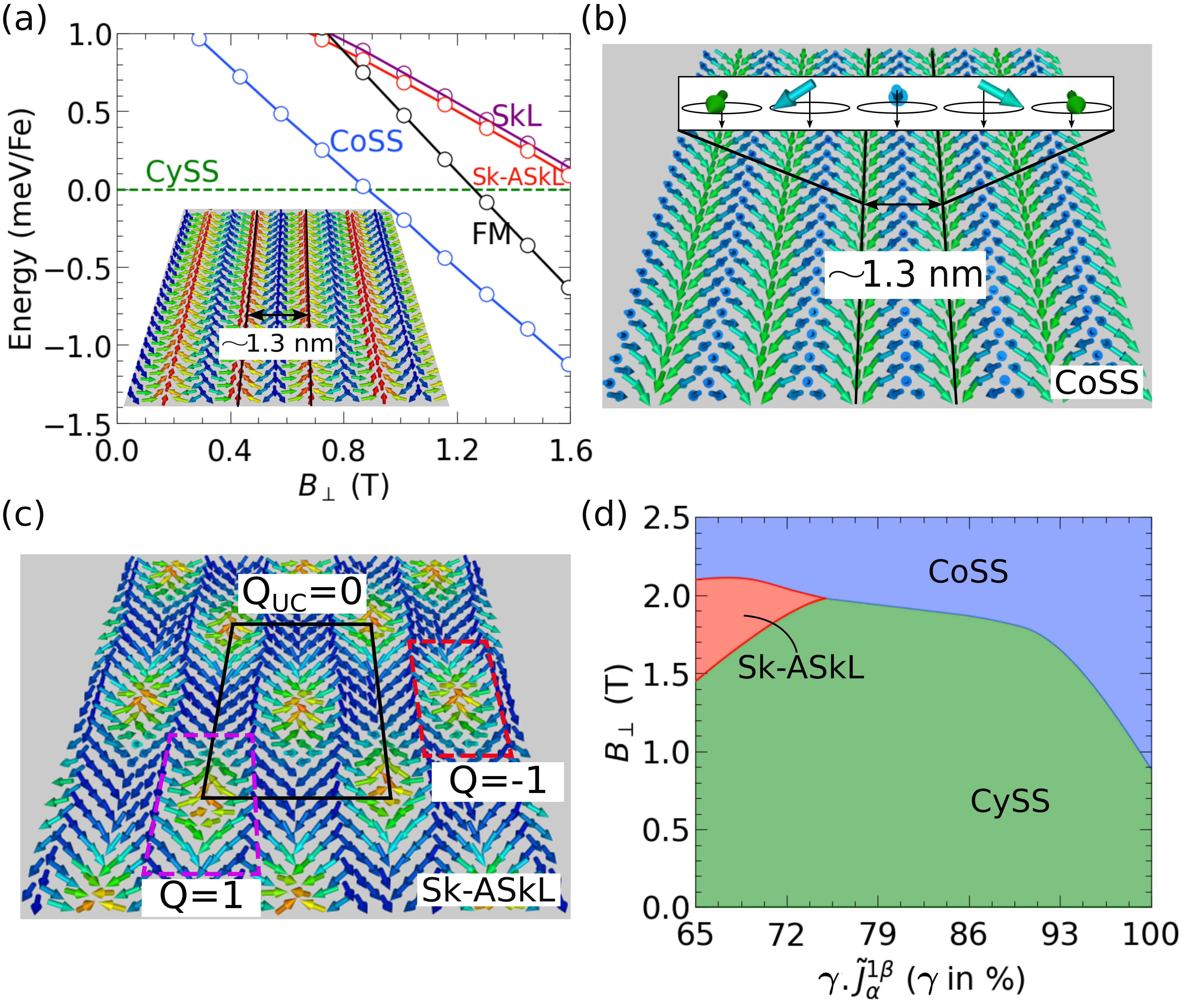}
    \caption{Zero temperature magnetic phases of \textit{pristine} 2Fe/GaAs(110) in the presence of magnetic field ($B_\perp$). (a) Total energies relative to the CySS state. The competing phases are CySS, CoSS, FM, SkL, and Sk-ASkL. Inset: Zero-field CySS ground state. (b) and (c) Spin textures of the CoSS and Sk-ASkL phases, respectively. (d) Phase stability vs. interlayer exchange ($\tilde{J}^{1\beta}_\alpha$). Reducing the nearest-neighbor exchange parameter by $\sim 25\%$, shifts the transition sequence to CySS$\rightarrow$Sk-ASkL ($Q_\textrm{UC}=0$)$\rightarrow$CoSS$\rightarrow$FM.   
    }
    \label{fig_3}
\end{figure}
Summarizing the material-specific parametrization of Eq.~\eqref{spin_hamiltonian}, it is evident that both sets of interaction parameters display significant anisotropy (heterogeneous behavior within a given shell) and frustration.
We then employ large-scale spin-lattice simulations, utilizing the SPIRIT code~\cite{SPIRIT}, to investigate the resulting magnetic phase behavior.
Specifically, by incorporating interaction parameters that extend beyond conventional short-range models, like nearest- and next-nearest-neighbor approximations, we systematically describe diverse magnetic phases and their stability below.

\noindent \textcolor{blue}{\textit{Tailoring magnetic phases.-}} 
The interplay and stability of competing magnetic phases in the \textit{pristine} 2Fe/GaAs(110) system are elucidated in Fig.~\ref{fig_3}. 
To begin this investigation, let the zero-field magnetic ground state of the thin Fe layer be precisely resolved using our spin-lattice simulations based on the model in Eq.~\eqref{spin_hamiltonian}. 
By specifically setting the iDMI to zero, exchange frustration has led to the formation of a SS state with a periodicity of $\lambda \approx 1.5$ nm (see Fig.~S3 in the SM~\cite{supp}). 
Such nanometer-scale SS solution is commonly found in 2D interfacial frustrated magnets~\cite{Bode_Nature2007, Dupe_NatCommun2014, FerrianiPRL, Nandy_PRL2016}.
In line with expectations, the period $\lambda$ is observed to shorten when chiral interaction ($i.e.$, the iDMI) is introduced into the model.
Consequently, the zero-field equilibrium state is a left-rotating CySS with $\lambda \approx$ 1.3 nm, see the inset of Fig.~\ref{fig_3}(a) and SM Fig.~S3(b).  
Additionally, a CoSS with identical periodicity exists as the nearest metastable phase, lying 1.4 meV/Fe higher in energy at zero magnetic field.

\begin{figure}
    \centering \includegraphics[width=0.95\linewidth]{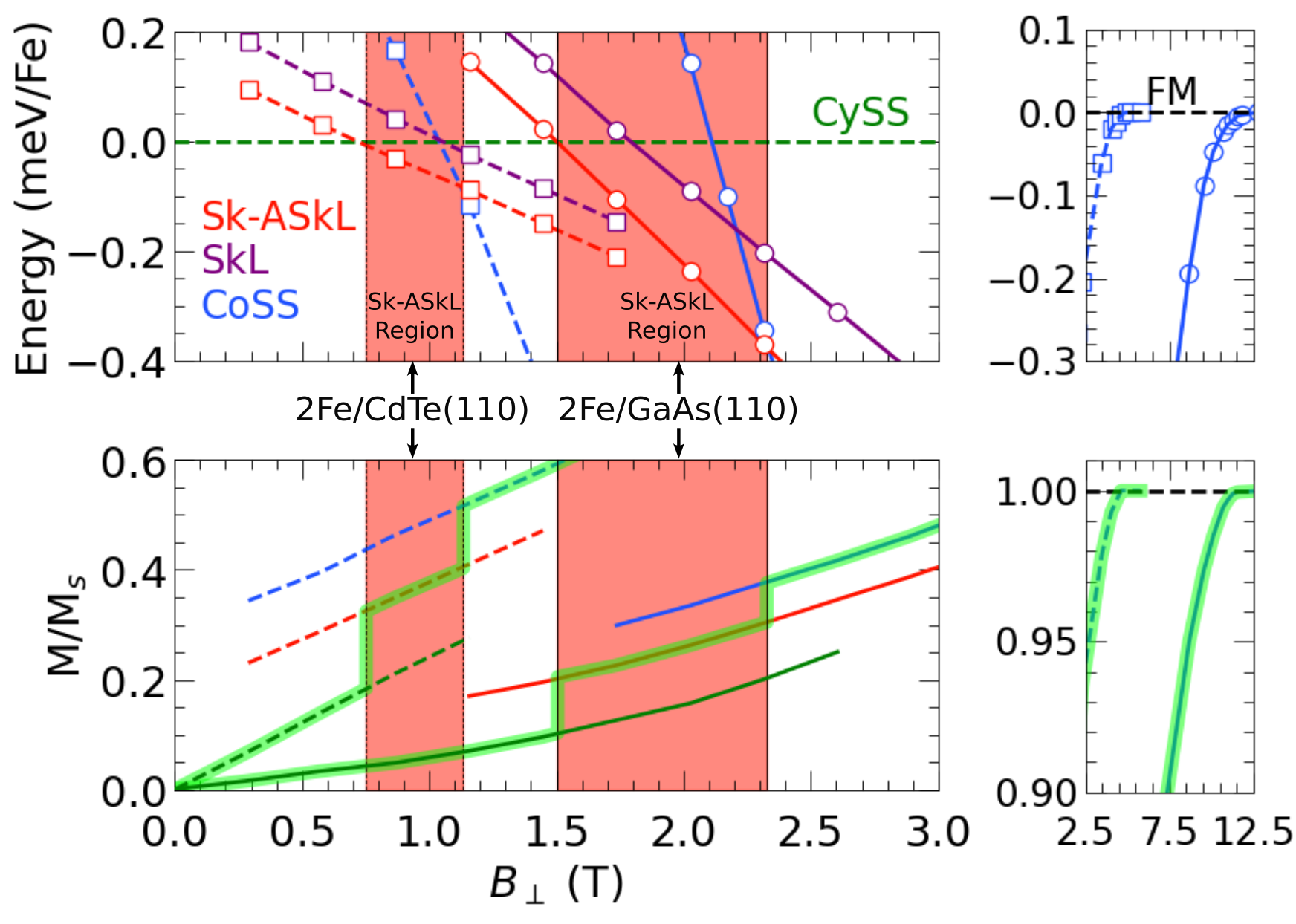}
    \caption{Stability of competing magnetic phases{--}CySS, Sk-ASkL, SkL, and FM{--}under $B_\perp$. (a) Relative total energy with respect to the CySS energy. Systems are strained 2Fe/GaAs(110) (solid lines) and \textit{pristine} 2Fe/CdTe(110) (dashed lines). The red-shaded region's vertical borders indicate critical $B_\perp$ fields for phase transitions. Lower panel: Magnetization discontinuities at critical fields ($M/M_\textrm{s}$, where $M_\textrm{s}$= 2.37 $\mu_\textrm{B}$ and 2.70 $\mu_\textrm{B}$ for 2Fe/GaAs(110) and 2Fe/CdTe(110), respectively). The transition between CoSS and FM phases is marked by the coalescence of energy and magnetization curves at high $B_\perp$ (rightmost figures).}
    \label{fig_4}
\end{figure}
 
Following Fig.~\ref{fig_3}(a), we investigate the energy variation of all stable phases in the presence of the Zeeman term $\mathcal{H}^Z$, generated by $B_\perp$ applied normal to the film. 
The zero energy level is set by the CySS state energy. 
The energetic intersection of the CySS and CoSS states, observed at a critical perpendicular magnetic field of $B_\perp \approx 0.9$ T, dictates a magnetic phase transition. 
The CoSS phase persists until a second critical field of 9 T, where the CoSS and FM phase lines coalesce.
The spin configuration of the CoSS is shown in Fig.~\ref{fig_3}(b). 
The later transition is characterized by a gradual and continuous increase in the normalized magnetization, $M/M_\textrm{S}$, ultimately reaching unity at the critical field.
In this pristine system, we find metastable Sk-ASkL and SkL phases, with the Sk-ASkL phase having slightly lower energy.
The spin configuration of Sk-ASkL, as shown in Fig.~\ref{fig_3}(c), is the \textit{net-zero} TC lattice carrying $Q_\textrm{UC}$=0.

To validate the impact of anisotropy, let us now consider 2Fe/GaAs(110), where directional disparities in exchange and iDMI are eliminated by averaging across each coordination shell. However, the DM vector anisotropy, a characteristic of the rectangular unit cell, persists~\cite{Hoffmann-NatCommun2017}. Here, in contrast to the $C_{2v}$ symmetry, the superposition of DM vectors from L1 and L2 yields an effective $C_{nv}$-type iDMI, instead of a $D_{2d}$-type iDMI.
This simplified model with isotropic exchange reveals a competition between the Sk-ASkL and SkL, with the former being energetically favored. 
Isotropic chiral magnets with $C_{4v}$ and $C_{3v}$ symmetry, on the other hand, invariably favor the SkL phase.
Exchange anisotropy, introduced by scaling the exchange parameters in the right half of each shell (inter- and intralayer) by 0.6, conclusively establishes the Sk-ASkL phase as the ground state, situated between the CySS and CoSS phases.
Details are provided in the SM Sec.~III~\cite{supp}.
Additionally, we employ a generic minimal model featuring two competing exchange constants and a nearest-neighbor DM vector to demonstrate that all observed magnetic phases are direct manifestations of the anisotropic interaction space, see SM Sec.~IV~\cite{supp}. This can stem from the low interfacial symmetry in our materials.

Despite the presence of exchange anisotropy in the \textit{pristine} 2Fe/GaAs(110) sample, the dominant FM exchange, particularly the interlayer coupling, precludes any topological lattice phase as the field-induced ground state. Interestingly, the reduction of interlayer $\tilde{J}$ by $\sim$ 25 \%, as evidenced in Fig.~\ref{fig_3}(d), enables the stabilization of the Sk-ASkL phase in the system. Thus, we have created a strained geometry by applying a compressive strain via constraining the lattice constant to a value 4\% below that of bulk GaAs (5.43 \AA).
The dashed lines in Fig.~S2 and its inset in the SM~\cite{supp} depict the calculated exchange and dominant DM vectors. 
A notable feature is the significant suppression of interlayer FM coupling $\tilde{J}$, attributed to the increased interlayer separation under strain.
As shown in Fig.~\ref{fig_4} (solid lines), the Sk-ASkL configuration persists as the equilibrium ground state for the strained 2Fe/GaAs(110) system within the field range 1.5{--}2.32 T.
In particular, the field-driven phase sequence follows CySS $\rightarrow$ Sk-ASkL $\rightarrow$ CoSS $\rightarrow$ FM. 
In the lower panel, the equilibrium magnetization, normalized to its saturation value $M_\textrm{S}$, is presented. 
Abrupt changes in magnetization, indicative of first-order phase transitions, are clearly resolved at the phase boundaries. 
The rightmost panels reveal a continuous, field-induced ($B_\perp \approx 10.5$ T) transition from the CoSS to the FM state, marked by gradual energy and magnetization convergence.

Figure~\ref{fig_4} (dashed lines) further demonstrates analogous phase behavior in the II-VI substrate sample, 2Fe/CdTe(110). This chiral example, even in its \textit{pristine} condition, develops a stable Sk-ASkL phase in the low-field regime (field range 0.75{--}1.12 T), accompanied by abrupt magnetization jumps. 
Consistent with the III-V sample, the energy and magnetization profiles merge at critical fields near 4.5 T. 
A comprehensive analysis of the 2Fe/CdTe(110) system, including a video rendering of spontaneous nucleation of skyrmions and antiskyrmions in our computational framework, is detailed in the SM Sec. V~\cite{supp}.

Intriguingly, the stability of such a 
\textit{net-zero} quasiparticle within a single magnetic layer opens exciting avenues for studying its current-driven dynamics in racetrack geometries. A crucial advantage is its distinctive ability to cancel equal and opposite Magnus forces, facilitating remarkably linear motion. As demonstrably shown by our current-driven dynamics simulations in SM Sec. VI~\cite{supp}, this \textit{net-zero} quasiparticle indeed exhibits linear motion, without any observable transverse component.

\noindent \textcolor{blue}{\textit{Conclusion.-}}
In summary, we unveil a surprising, magnetic field-driven, thermodynamically stable Sk-ASkL phase in $C_{1v}$-symmetric interfacial magnets. 
This novel phase, with a \textit{net-zero} global topological charge ($Q_\textrm{UC}=0$), overcomes the expected annihilation of skyrmion-antiskyrmion pairs, stabilized by symmetry-enforced anisotropic interactions. 
Employing a combination of density functional theory and large-scale spin-lattice simulations, we demonstrate that the interplay of frustrated exchange and iDMI in Fe/GaAs(110) and Fe/CdTe(110) systems stabilizes this unconventional phase. 
Moreover, we identify a sequence of field-driven transitions: from the CySS phase to the Sk-ASkL phase, followed by a CoSS state, and finally a gradual merging between CoSS and FM states. 
This work highlights the crucial role of symmetry engineering in controlling unconventional topological magnetism.

\noindent \textcolor{blue}{\textit{Acknowledgments.-}} A.K.N. and S.B. acknowledge the support from the Department of Atomic Energy (DAE), Government of India, through the project Basic Research in Physical and Multidisciplinary Sciences via RIN4001. 
A.K.N. and S.B. acknowledge the computational resources, Kalinga cluster, at the National Institute of Science Education and Research (NISER), Bhubaneswar, India. 
A.K.N. thanks Prof. P. M. Oppeneer for the Swedish National Infrastructure for Computing (SNIC) facility. 
The authors thank Prof. Stefan Bl\"ugel and Dr. Nikolai S. Kiselev for fruitful discussions. 
A.K.N. thanks Prof. Amit Agarwal for critical reading of the manuscript and stimulating discussions.
\bibliographystyle{apsrev4-1}
\bibliography{Bibliography}{}
\clearpage
\newpage

\begin{onecolumngrid}
\vspace{-2em}

\begin{center}	
	{
		\fontsize{12}{12}
		\selectfont
		\textbf{Supplemental material for ``Skyrmion-Antiskyrmion Lattice: A \textit{Net-Zero} Topological Phase in Low-Symmetry Frustrated Chiral Magnets"\\[5mm]}
	}
	
	\normalsize Sayan Banik\orcidA{}$^{1}$ and Ashis K. Nandy\orcidB{}$^{1}$\\
	{\small $^1$\textit{School of Physical Sciences, National Institute of Science Education and Research,\\ 
			An OCC of Homi Bhabha National Institute, Jatni 752050, India}\\[0.5mm]}
	
\end{center}	

\section{Methodology}\label{Sec:S1}
\renewcommand{\thetable}{\textbf{S\arabic{table}}}
\renewcommand{\thefigure}{\textbf{S\arabic{figure}}}
\subsection{\textit{Ab~initio} electronic structure calculations and material parameters determination}
We have employed spin-polarized density functional theory (DFT) calculations, as implemented in the Vienna Ab-initio Simulation Package (VASP)~\cite{hafner2008ab,kresse1996,Kresse}, to investigate the electronic and magnetic properties of transition-metal/semiconductor (TM/SC) heterostructures.
For both the 2Fe/GaAs(110) and 2Fe/CdTe(110) systems, the asymmetric slab geometries consist of an Fe bilayer on a 9-layer substrate. This number of substrate layers is sufficient to yield consistent results with converged magnetic properties.
The dimensions of the surface unit cells are determined using the experimental bulk lattice constants of the GaAs (III-V) and CdTe (II-VI) semiconductors, which are 5.565 \AA~and 6.478 \AA, respectively. 
Vacuum regions, approximately 12~\AA~thick, are added above and below the slab along the growth direction ($z$).
Our DFT calculations are performed using a plane-wave projector-augmented-wave (PAW) implementation~\cite{paw1, paw2}. The local density approximation (LDA) with the Vosko-Wilk-Nusair (VWN) functional~\cite{vwn80} has been used for the exchange-correlation potential.
In the self-consistent calculations, integration over the two-dimensional (2D) Brillouin zone (BZ) has been carried out using a 16$\times$16$\times$1 $\Gamma$-centered $k$-point mesh.
The calculations use plane wave basis states with a cutoff energy of 500 eV.
Atomic positions in the slabs are relaxed until the forces on all atoms in the magnetic Fe layers and the adjacent three substrate layers (near the Fe/semiconductor interface) are reduced to below 0.005 eV/\AA.
It is important to note that relaxing substrate layers deeper than the fourth layer did not significantly alter the final results.
The self-consistent total energy calculations converged to an accuracy of 10$^{-7}~\rm {eV}$.
The relaxed geometry of the system is accurately represented in Fig.~\ref{fig_1}.
To quantify the uniform biaxial strain in the 2Fe/GaAs(110) system, the strain percentage was calculated relative to the experimental lattice constant using strain = $\frac{a-a_\textrm{expt}}{a_\textrm{expt}}\!\times\!100 \%$.
Therefore, in the case of compression, the sign of strain is negative. 

The magnetic properties and magnetic interactions of relaxed 2Fe/GaAs(110) and 2Fe/CdTe(110) slabs are determined using the all-electron full-potential scalar-relativistic Korringa-Kohn-Rostoker (KKR) Green function method~\cite{Bauer_thesis, N_Papanikolaou_KKR2002}, with spin-orbit coupling included self-consistently as implemented within the JuKKR code~\cite{KKR-Code_2022}.
This approach employs an exact description of atomic cells and is rooted in multiple-scattering theory~\cite{sp_kkr}. 
The relaxed slab geometry has an equal vacuum layer thickness of 12~\AA~on both sides.
For consistency, the exchange-correlation interactions are treated within the local density approximation (LDA), as formulated by VWN functional~\cite{vwn80}.
The effective potentials and fields were computed within the atomic sphere approximation (ASA), employing an angular momentum cutoff of $l_{max}=3$. 
The energy contour has 38 complex energy points in the upper complex plane, and it incorporates 38 Matsubara poles.
A Fermi smearing parameter corresponding to 473 K is used. 
A 40$\times$40$\times$1 mesh is utilized for $k$-space integrations over the 2D BZ in the self-consistent calculations.

The infinitesimal rotation method~\cite{LIECHTENSTEIN198765, Rot2}, implemented within a generalized relativistic framework~\cite{Noncol1,NonCol2} is used to determine the Heisenberg exchange and the Dzyaloshinskii-Moriya (DM) vectors with a finer $k$-points grid mesh of 120$\times$120$\times$1.
To determine uniaxial magnetocrystalline anisotropy (MCA), the converged potential was utilized to calculate the total band energy ($E_i,~i\in$ {x, y, z})  for each magnetization orientation ($x, y$ and $z$), with the same $k$-point mesh.
The minimum energy difference $E_x-E_z$ or $E_y-E_z$ defines the magnitude of the MCA, K. The negative (positive) value of K refers to in-plane (out-of-plane) MCA.

\subsection{Numerical calculations for various magnetic phases: Atomistic Spin Dynamics simulations}
Numerical solutions of the extended Heisenberg Hamiltonian (Eq.~(1) in the Main Text), using magnetic interaction parameters extracted from DFT, are employed to determine the various magnetic phases of the 2Fe/GaAs(110) and 2Fe/CdTe(110) systems.
This involves large-scale atomistic Monte Carlo (MC) simulations~\cite{MC_1,sim_ann}, implemented in the Spirit code~\cite{SPIRIT}, a specialized platform for atomistic spin dynamics modeling that provides detailed, atom-resolved magnetic configurations.
In our simulations, we have considered interactions extending up to 12 intralayer and interlayer shells around each Fe atom.
Beyond this, the parameters are negligibly small, and further neighboring shells do not alter the computed spin configuration.   
This invariance demonstrates that the dominant magnetic interactions,  including frustration, are adequately represented by our initial parameterization.  

Our simulation cell is a bilayer structure with dimensions of $(80 \times 80) \times 2$ atomic sites.
Simulated annealing~\cite{sim_ann} protocols are employed, whereby the system begins with a random spin state at $T_0 = 300$~K, and is gradually cooled to $T_f = 10^{-5}$~K (effectively 0~K).
Starting from a random spin configuration, we intentionally apply open boundary conditions (OBC) to facilitate the spontaneous formation of the spin-spiral (SS) state, which is expected to generate an SS state with a period ($\lambda$) close to its optimized value.
To establish the equilibrium $\lambda$ value, we further employ constrained initialization protocols. 
This involves subjecting preconfigured SS states, each with a different $\lambda$ value, to energy relaxation via thermal cycling.
Here, periodic boundary conditions (PBC) are applied in the 2D domain.
In each $\lambda$ value, the system has been initialized at a temperature of 20 K and then cooled to 0 K in steps of 0.5 K.
A comparative analysis of the free energies of the final configurations identifies the thermodynamically stable period. 

In the presence of the Zeeman energy term in our model, we initialize the spin configuration from both a random orientation and an SS state. 
The external magnetic field $\mathbf{B}_{ext}$ is applied along the $z$-direction. 
We also allow the system to relax with OBC to confirm the spontaneous nucleation of skyrmions (Sks) and antiskyrmions (ASks).

\begin{figure}[H]
    \centering
    \includegraphics[width=0.8\linewidth]{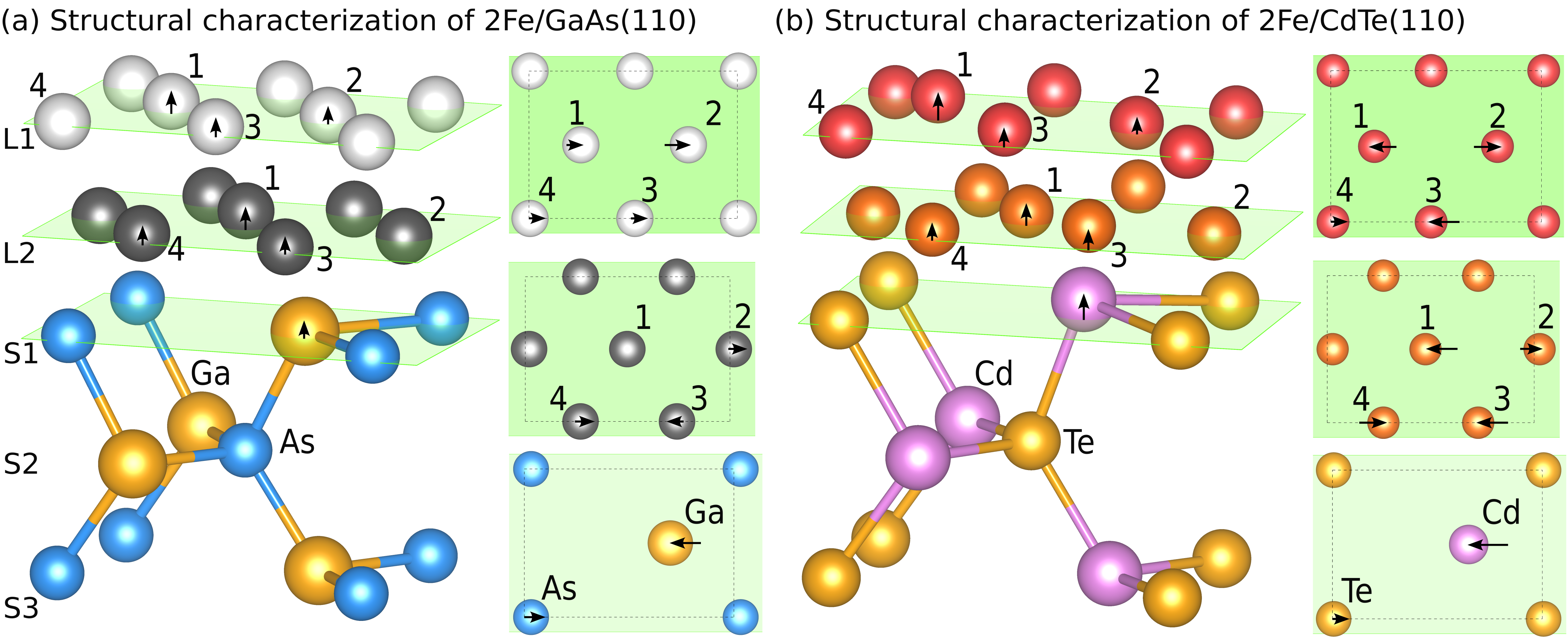}
    \caption{The relaxed atomic geometries near the interface are shown for (a) the 2Fe/GaAs(110) heterostructure and (b) the 2Fe/CdTe(110) heterostructure. The displacement of the atoms within the plane is indicated by the in-plane arrows. The out-of-plane arrows depict the outward displacement of the atoms from the atoms positioned at the lowest point in the corresponding layer. For example, in L1 and L2, the lowest positions are Fe-4 and Fe-2 atoms, respectively. The length of the arrows is proportional to the magnitude of the displacement.}
    \label{fig_1}
\end{figure}
\section{Interface characterization of 2F\lowercase{e}/G\lowercase{a}A\lowercase{s}(110) and 2F\lowercase{e}/C\lowercase{d}T\lowercase{e}(110) systems}

Figures~\ref{fig_1}(a) and (b) show the relaxed interface structures of 2Fe/GaAs(110) and 2Fe/CdTe(110), respectively. 
The rectangular unit cell is defined by the 2D lattice vectors ($a$, 0, 0), (0, $a/\sqrt{2}$, 0), where `$a$' represents the lattice constant of the bulk semiconductor. 
For zincblende binary semiconductors, the ideal interlayer distance between (110) planes is $a / (2\sqrt{2})$. 
In the heterostructure, this spacing is preserved in the substrate layers after the initial five layers (the first two Fe layers, L1 and L2, and the first three substrate layers, S1, S2, and S3).
Relaxed interlayer distances are quantified in units of this ideal distance.
The relaxed positions of the first five layers are presented in Tab.~\ref{tab:atomic_positions}, while Fig.~\ref{fig_1} uses arrows to visualize the atomic displacements. 
In this coordinate system, $z = 0$ corresponds to the position of the fourth semiconductor layer from the interface. 
In-plane coordinates are given in units of the lattice vectors. 
In these $C_{1v}$ systems, the substrate-induced breaking of left-right symmetry results in atomic movement within the plane along the $x$-direction, leaving the $y$-components of the atomic positions unchanged.
The non-uniform height ($z$-components) within these layers indicates interface roughness, a characteristic attributed to the surface corrugation commonly observed in III-V and II-VI semiconductor surfaces.

\begin{table}[h!]
    \centering
    \renewcommand{\thetable}{\textbf{S\arabic{table}}}
    \begin{tabular}{|c|c|c|c|c|c|c|c|}
        \hline
        & & \multicolumn{3}{c|}{2Fe/GaAs(110)} & \multicolumn{3}{c|}{2Fe/CdTe(110)} \\ \cline{3-8}
        Layers & Atoms & x (ideal) & y (ideal) & z (ideal) & x (ideal) & y (ideal) & z (ideal) \\ \hline
        \multirow{4}{*}{L1}    
        & Fe-1 &0.26 (0.25) &0.5 (0.5) &5.158 (5) &0.21 (0.25) &0.5 (0.5) &4.879 (5) \\
        & Fe-2 &0.77 (0.75) &0.5 (0.5) &5.129 (5) &0.79 (0.75) &0.5 (0.5) &4.561 (5) \\
        & Fe-3 &0.51 (0.50) &0.0 (0.0) &5.148 (5) &0.45 (0.50) &0.0 (0.0) &4.811 (5) \\
        & Fe-4 &0.01 (0.00) &0.0 (0.0) &5.102 (5) &0.01 (0.00) &0.0 (0.0) &4.620 (5) \\
        \hline
        \multirow{4}{*}{L2}    
        & Fe-1 &0.50 (0.50) &0.5 (0.5) &4.219 (4) &0.45 (0.50) &0.5 (0.5) &4.250 (4) \\
        & Fe-2 &0.02 (0.00) &0.5 (0.5) &4.087 (4) &0.08 (0.00) &0.5 (0.5) &3.884 (4) \\
        & Fe-3 &0.74 (0.75) &0.0 (0.0) &4.110 (4) &0.70 (0.75) &0.0 (0.0) &4.060 (4) \\
        & Fe-4 &0.27 (0.25) &0.0 (0.0) &4.144 (4) &0.29 (0.25) &0.0 (0.0) &3.952 (4) \\
        \hline
        \multirow{2}{*}{S1}    
        & Ga/Cd &0.70 (0.75) &0.5 (0.5) &3.178 (3) &0.65 (0.75) &0.5 (0.5) &3.278 (3) \\
        & As/Te &0.03 (0.00) &0.0 (0.0) &3.152 (3) &0.01 (0.00) &0.0 (0.0) &3.163 (3) \\
        \hline
        \multirow{2}{*}{S2}    
        & Ga/Cd &0.25 (0.25) &0.0 (0.0) &2.065 (2) &0.25 (0.25) &0.0 (0.0) &2.067 (2) \\
        & As/Te &0.51 (0.50) &0.5 (0.5) &2.065 (2) &0.51 (0.50) &0.5 (0.5) &2.088 (2) \\
        \hline
        \multirow{2}{*}{S3}    
        & Ga/Cd &0.75 (0.75) &0.5 (0.5) &1.028 (1) &0.75 (0.75) &0.5 (0.5) &1.033 (1) \\
        & As/Te &0.00 (0.00) &0.0 (0.0) &1.038 (1) &0.00 (0.00) &0.0 (0.0) &1.038 (1) \\
        \hline
    \end{tabular}
    \caption{After structural relaxation, atomic positions are provided for 2Fe/GaAs(110) and 2Fe/CdTe(110). Ideal positions are tabulated in parentheses.}
    \label{tab:atomic_positions}
\end{table}

\section{Results for 2F\lowercase{e}/G\lowercase{a}A\lowercase{s}(110) from KKR calculations}
\subsection{Atomistic magnetic interaction parameters}

Table~\ref{MagMom_GaAs} tabulates the \textit{ab initio} calculated magnetic moments of each Fe atom. 
Within each layer, four Fe atoms are numbered as shown in Figs.~\ref{fig_1}(a) and (b).
Fe moments in the top L1 layer are consistently near 2.5 $\mu_\textrm{B}$. In contrast, the second layer displays a variable magnetic moment influenced by the chemical environment in the substrate.
For all atomistic spin-lattice simulations performed with the SPIRIT code, we employ a simplifying assumption of homogeneous magnetic moments, using an average value of 2.37 $\mu_\textrm{B}$ per Fe atom. This simplification is made without loss of generality.
The calculated MCA is out-of-plane.

\begin{table}[h]
	\centering
	\begin{tabular}{|c|c|c|c|c|}\hline
	Magnetic layers&Fe atom& Magnetic mom. in $\mu_B$&Average ($\mu_B$)&K (meV/Fe atom) \\ \hline
    \multirow{4}{*}{L1}
        &Fe 1&2.54&& \\ \cline{2-3}
        &Fe 2&2.53&&  \\ \cline{2-3}
        &Fe 3&2.49&&  \\ \cline{2-3}
        &Fe 4&2.51&2.37&0.11  \\ \cline{1-3}
    \multirow{4}{*}{L2}
        &Fe 1&2.03&&  \\ \cline{2-3}
        &Fe 2&2.40&&  \\ \cline{2-3}
        &Fe 3&2.32&&  \\ \cline{2-3}
        &Fe 4&2.16&&  \\ \hline
        \end{tabular}
	\caption{Magnetic moments of Fe atoms in the magnetic layer and the MCA parameter.}
	\label{MagMom_GaAs}
\end{table}

Figure~\ref{GaAs_Int} shows the interaction parameters (exchange and DM vectors) for each atom, as determined from KKR calculations, for both the pristine and strained Fe/GaAs(110) configurations.
The inhomogeneous chemical environment within these low-symmetry heterostructures leads to significant variations in the exchange and DM vectors. 
This behavior distinguishes them as a different class of 2D magnets, in contrast to transition-metal/heavy-metal heterostructures.
The eight Fe atoms exhibit a range of exchange coupling strengths and varying degrees of frustration with their neighboring atoms.
For example, the maximum exchange strength is 28.1 meV within a layer and 37.7 meV between layers, associated between atom 1 and atom 3 of layer L2 and atom 4 of L1 and L2, respectively. 
The inset provides a detailed view of the intralayer and interlayer DM vectors, specifically for the first two neighboring shells. 
Furthermore, the inhomogeneity in exchange parameters within a given shell is visually represented by colored spheres.  
The length of the arrows is proportional to the strength of the DM vectors.
The DM vectors are left- and right-rotating in the L1 and L2 layers, respectively, while the interlayer DM vectors are right-rotating.
Compressive strain significantly reduces the interlayer exchange interactions for all atoms, especially in the first shells. 
This reduction is attributed to the increased interlayer distance resulting from the strain. 
The rotational sense of the DM vectors (both intra- and interlayer) is unchanged with the application of strain. 
\begin{figure}[h!]
    \centering
    \includegraphics[width=0.8\linewidth]{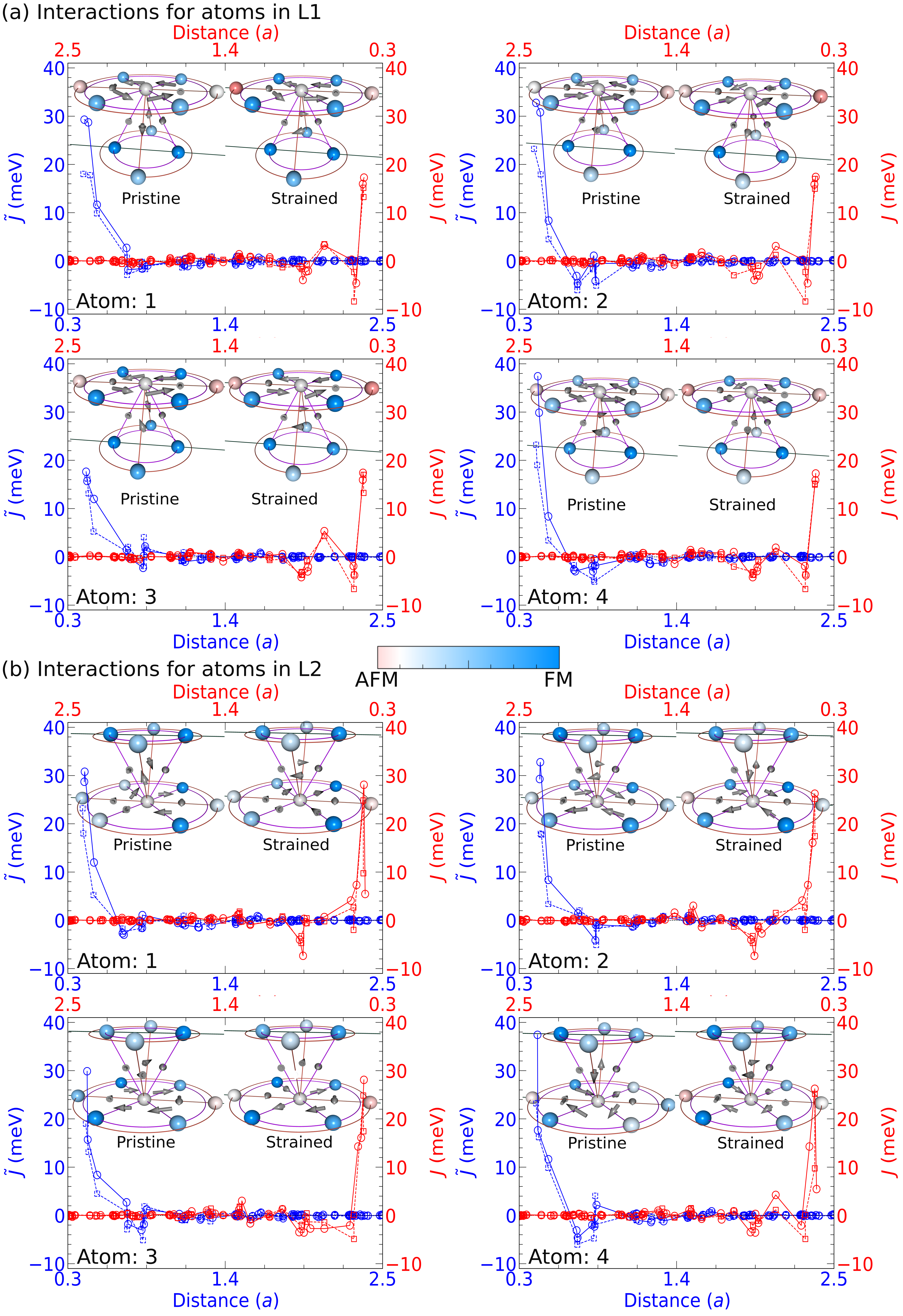}
    \caption{Magnetic interactions of 2Fe/GaAs(110): Magnetic interactions (a) for the atoms in L1 and (b) for the atoms in L2. Exchange interactions are shown for the pristine (solid lines) and strained (dashed lines) systems. The inset depicts DM vectors for both. Spheres within the first two neighboring shells are color-coded by exchange coupling strength (saturation indicates strength) and type (blue: ferromagnetic (FM), red: antiferromagnetic (AFM) between Fe atoms).}
    \label{GaAs_Int}
\end{figure}
\subsection{Exchange frustration driven spin spiral}
The significant exchange frustration in the system, evident in Fig.~\ref{GaAs_Int}, results in an SS solution, designated as a frustrated SS. 
In this case, we have simplified the full Hamiltonian (Eq.~(1) in the main text) in MC simulations by eliminating secondary interaction terms such as DM interaction (DMI), MCA, and Zeeman interaction. 
Our simulations demonstrate the spontaneous emergence of a cycloidal spin spiral (CySS) with an approximate periodicity of 1.5 nm (Fig. \ref{DMI_Effect}(a)), thus indicating that frustrated exchange interactions can induce SS order. However, the chiral nature of this state only becomes apparent when DM vectors are included in the Hamiltonian. 
Considering the full Hamiltonian, including exchange, DMI, and out-of-plane MCA, the spiral wavelength is slightly reduced. This yields a chiral state (left-rotating CySS) with an approximate period of 1.3 nm, see Fig.~\ref{DMI_Effect}(b).

\begin{figure}[H]
    \centering
    \includegraphics[width=0.85\linewidth]{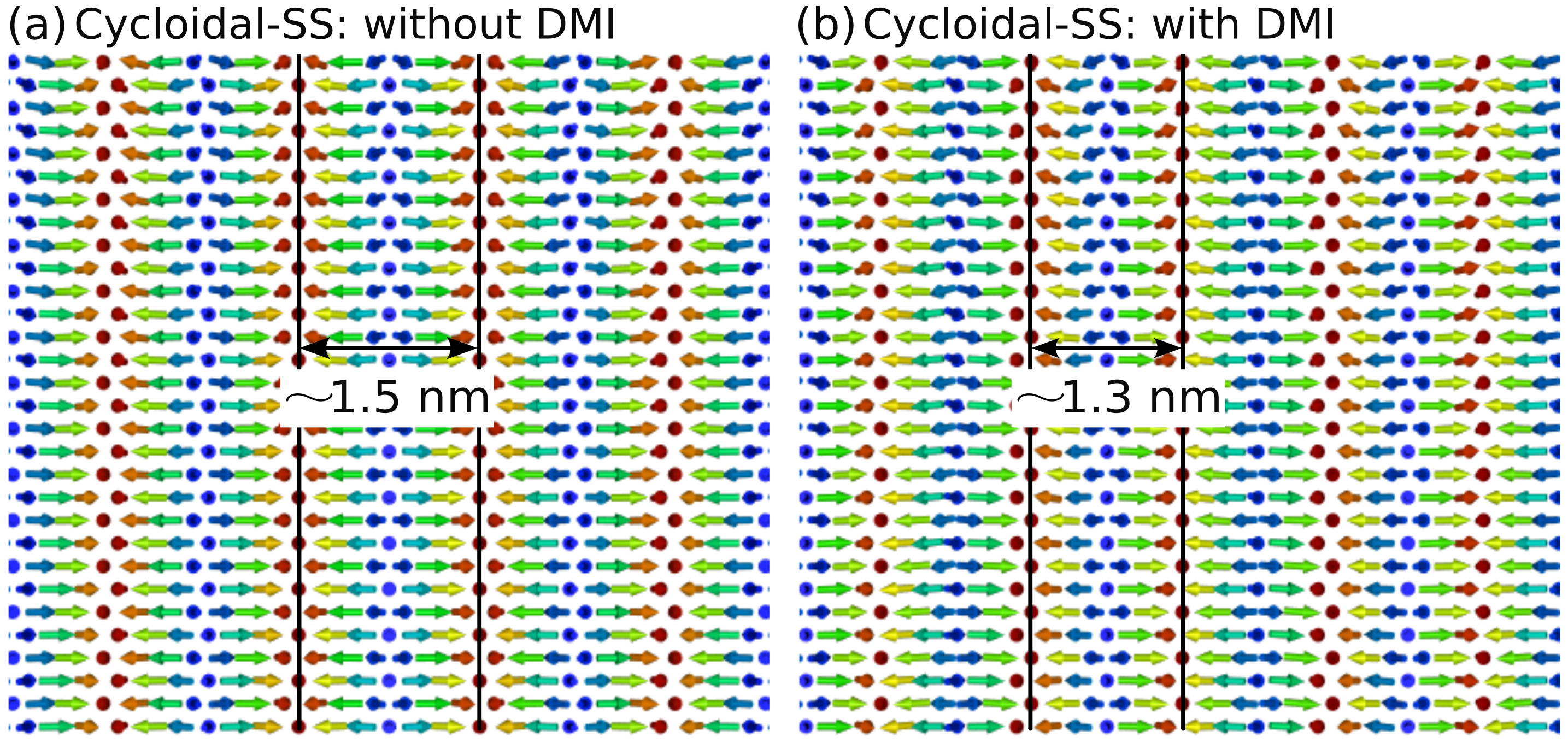}
    \caption{The top view of the SS states. For clarity, only a single Fe layer is depicted. (a) The frustrated SS state, which arises spontaneously due to exchange frustration. (b) The left-handed CySS.}
    \label{DMI_Effect}
\end{figure}
\vspace{-15em}
\subsection{Results with average spin model: Role of symmetry induced anisotropy in interaction parameters}
Here, we explain how systems with $C_{1v}$ symmetry differ from conventional isotropic chiral magnets with higher symmetry, such as $C_{3v}$ and $C_{4v}$. 
Ultrathin transition-metal films on heavy-metal substrates are well-suited for realizing these higher symmetries, where the hexagonal skyrmion lattice (SkL) is a common equilibrium phase. 
A key characteristic of these systems is that they can be described by a single-atom-per-layer unit cell, leading to isotropic interaction parameters. 
For instance, the DM vectors have equal magnitudes and are orthogonal to the bond connecting two magnetic atoms, while the exchange parameters are uniform within the neighboring shell.
In these systems, a chiral CySS state undergoes a magnetic field-induced phase transition to the SkL, accompanied by a discontinuous jump in magnetization.

\begin{figure}[H]
    \centering
    \includegraphics[width=0.7\linewidth]{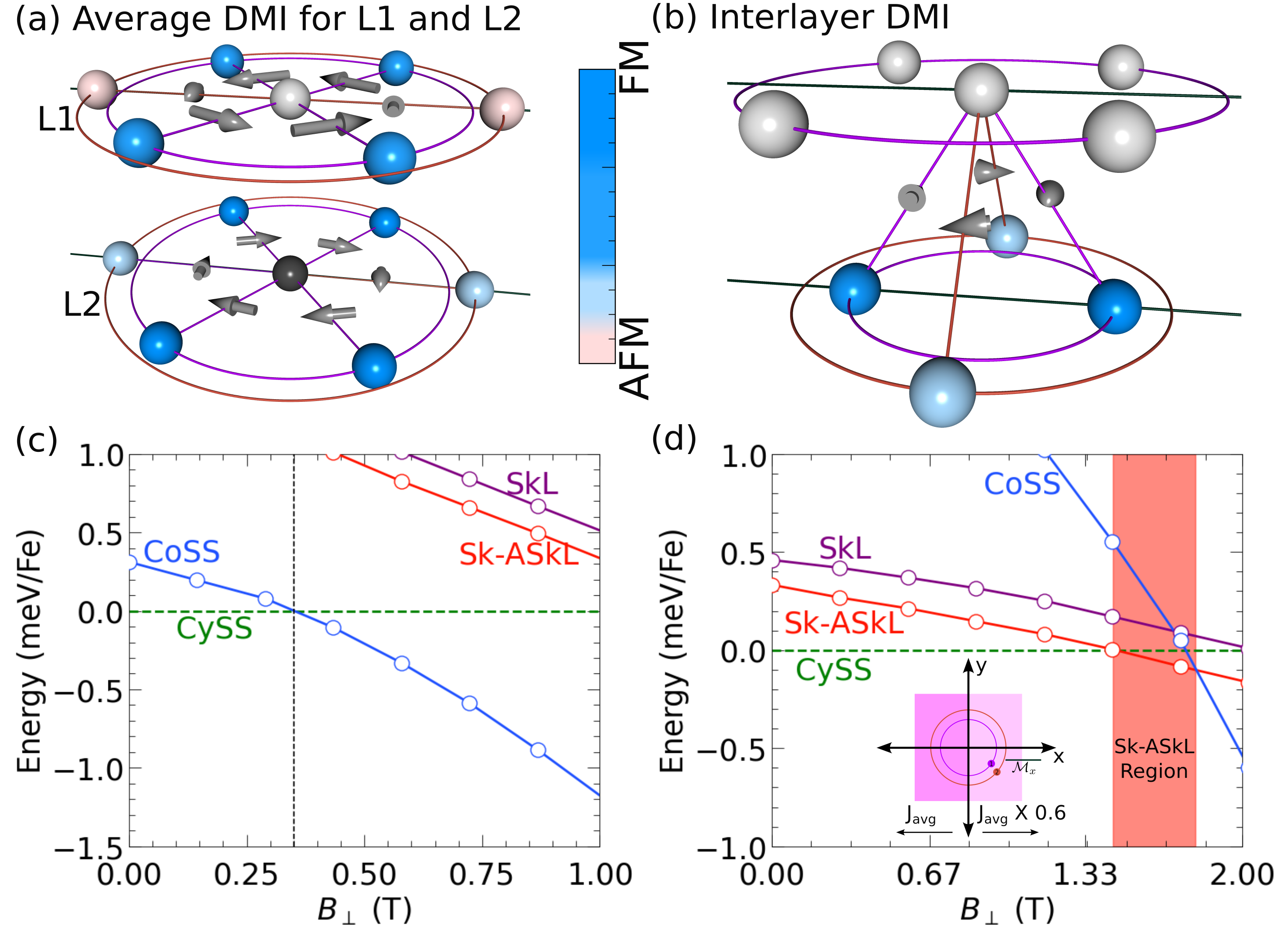}
    \caption{Results with average interactions for 2Fe/GaAs(110) system. (a) Intralayer DMI averaged over L1 and L2, showing left- and right-rotating behavior in L1 and L2, respectively. The exchange interaction strengths are color-coded: Blue for FM and red for AFM. (b) The effective sum of intralayer DMIs in L1 and L2. As the left-rotating DMI in L1 is greater in magnitude than the right-rotating DMI in L2, their effective sum is a left-rotating DMI with reduced magnitude. (c) Right-rotating average interlayer DMI selects CySS chirality. 
    (d–f) Energy diagrams for the pristine average system (d), 30\%-reduced $\tilde{J}^{1\beta}_\alpha$ on that system (e), and further asymmetric 40\% reduction on atoms with positive $x$ coordinates (f).}
    \label{Avg_Res}
\end{figure}

Due to the reduced $C_{1v}$ symmetry, our systems have a more complex parameter space, as reflected in the anisotropy of exchange and DMI observed in the material-specific parameters. Therefore, it is important to address the role of the semiconductor substrate, especially for magnetic films on (110) surfaces of III-V and II-VI semiconductors. 
To this end, we utilize an average spin model, which we subsequently modify to account for the anisotropic parameters, thereby stabilizing the Sk-ASkL as the equilibrium phase.
This approach eliminates the inhomogeneity (anisotropy) in the exchange interaction and DMI magnitudes. 
This simplification allows us to describe the system with parameters closer to those of an isotropic model, particularly with $C_{2v}$ symmetry.
The exchange parameters for each inequivalent Fe atom are averaged across each neighboring coordination shell. 
Further averaging over all inequivalent sites yields a uniform exchange interaction scheme.
For instance, the nearest-neighbor and next-nearest-neighbor intralayer exchange interactions are 16 meV (19 meV) and $-2.6$ meV (5.9 meV) for layer L1 (L2), respectively. 
The corresponding interlayer coupling values are 27.9 meV and 10.1 meV.  
Within this approach, the exchange interaction is uniform within a given shell, making all Fe atoms equivalent in their intralayer and interlayer couplings. 
The colored balls surrounding a representative central Fe atom in Fig.~\ref{Avg_Res}(a) and (b) illustrate this uniform exchange interaction scheme for the first two neighboring shells. 
Subsequently, the averaging procedure is also applied to the components of the DM vectors.
In these figures, intralayer and interlayer DM vectors are represented by arrows, where uniform arrow length within a shell indicates the same magnitude.
The DM vectors are all in-plane, with their rotation being left-handed within the L1 layer, right-handed within the L2 layer, and left-handed for interlayer interactions.
The corresponding magnitudes of the nearest-neighbor DMI are 1.12, 0.6, and 0.1 meV. Unlike the high-symmetry isotropic cases, the non-orthogonality of the DM vectors to the bond becomes apparent due to the absence of mirror planes parallel to those bonds, see nearest-neighbor arrows in Fig.~\ref{Avg_Res}(a).

Following the approach used by Hoffmann et al.~\cite{Hoffmann-NatCommun2017} for two-layer systems (possessing $C_{2v}$ interfacial symmetry), we calculate the effective sum of DM vectors from intralayer and interlayer contributions. 
Our calculation demonstrates that this summation results in a right-rotating effective DMI of $C_{nv}$-type. 
This differs from $C_{2v}$ systems, where the summation yields an effective $D_{2d}$-symmetric DM vectors. 
It is important to note that the ($D_{2d}$) symmetry stabilizes isolated ASk, whereas the magnetic ground state is a ferromagnet~\cite{Zimmermann_PRB2014}. 
In contrast, we observe a left-rotating CySS as the zero-field ground state. This state transitions to a conical spin spiral (CoSS) under an external perpendicular magnetic field, ($B_\perp$) (see Fig.~\ref{Avg_Res}(c)). 
The final high-field phase is the saturated ferromagnetic phase. 
Across a broad range of magnetic fields, the skyrmion-antiskyrmion lattice (Sk-ASkL) exhibits lower energy than the SkL, though their energies remain higher than that of the equilibrium phases. 
The lower energy of Sk-ASkL is attributed to frustrated isotropic exchange and anisotropic DM vector orientation. 
This DMI anisotropy is intrinsic, given that all parameters have been derived from calculations within a rectangular unit cell.
Notably, pure isotropic model parameters, with DM vectors orthogonal to the bonds, result in the SkL having lower energy than the Sk-ASkL.

Strong interlayer exchange coupling has been identified as one factor preventing any lattice phase from becoming the equilibrium phase. 
Furthermore, the energy difference between the SkL and the Sk-ASkL increases with decreasing interlayer exchange. 
Concurrently, the destabilization energy of Sk-ASkL decreases with respect to the equilibrium phases. 
To promote the Sk-ASkL as the equilibrium phase within a certain magnetic field range, we now reduce the interlayer exchange by 30\%. 
This reduction in 2Fe/GaAs(110) parameters has been achieved by applying compressive strain to increase the interlayer separation between Fe layers, thus reducing the interlayer exchange. 
An important feature absent in the average model is the left-right symmetry breaking, which is characteristic of these heterostructures. 
To account for this, we introduce a multiplicative factor to the exchange parameters, scaling those right of the $y$-axis by 0.6 (see the inset of Fig.~\ref{Avg_Res}(d)). 
Figure~\ref{Avg_Res}(d) clearly demonstrates the Sk-ASkL phase as the energetically most favorable state within the magnetic field range of 1.55 T to 1.72 T. 
Thus, our analyses, based on an average spin model with controlled anisotropy in parameters (characterizing a new class of chiral magnet), illustrate the key role that anisotropy plays in stabilizing the Sk-ASkL phase, particularly in materials with low symmetry.  
\section{Atomistic spin lattice model Hamiltonian}
\begin{figure}
    \centering
    \includegraphics[width=0.85\linewidth]{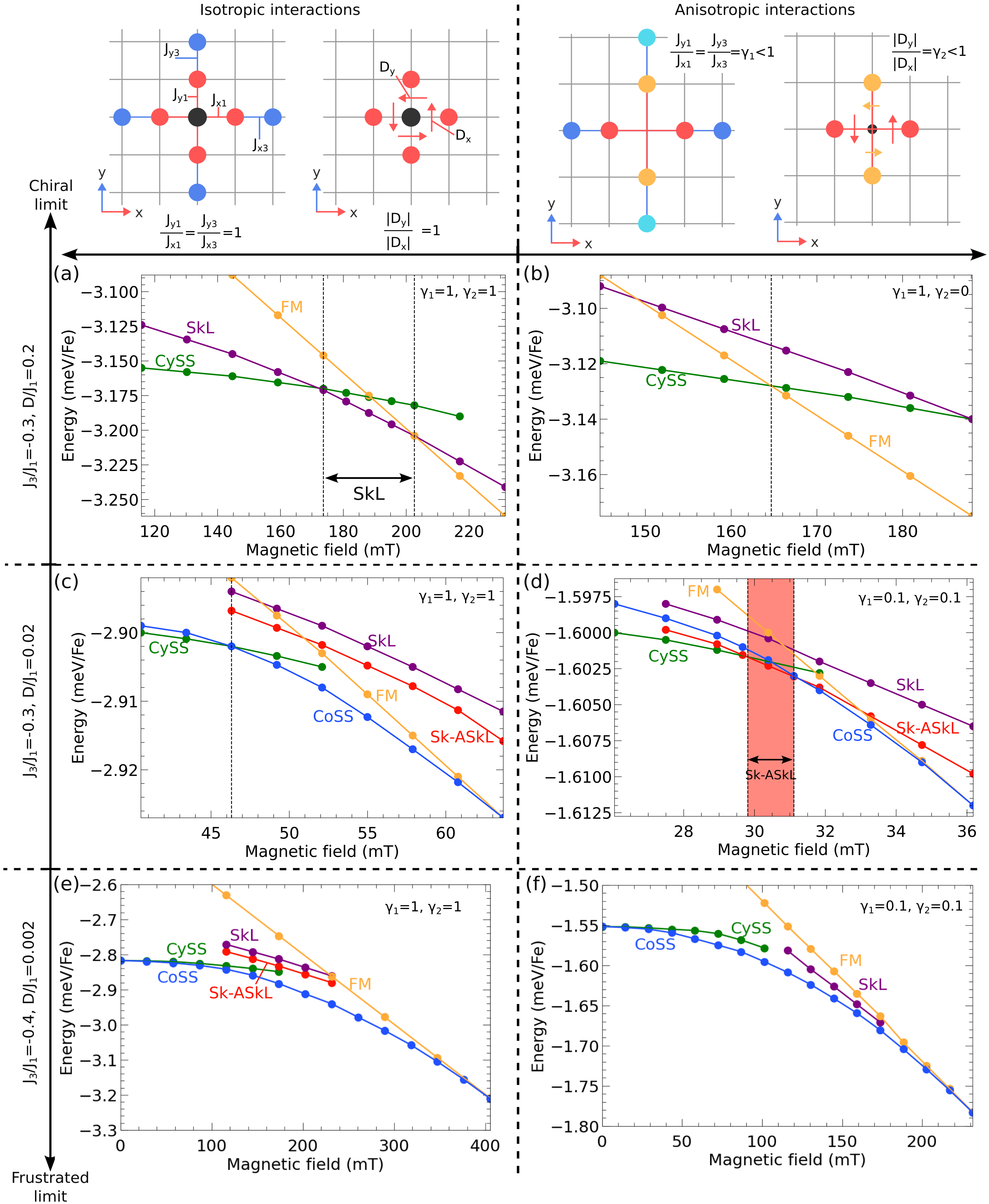}
    \caption{Energy line diagrams for the $J_1-J_3-D$ model. (a) and (b) Phase diagrams in the strong DMI limit for isotropic and anisotropic cases, respectively. (c) and (d) depict magnets where exchange frustration and DMI are delicately balanced, for isotropic and anisotropic cases, respectively. Notably, tailoring a Sk-ASkL phase becomes possible with the introduction of anisotropy in both the exchange interaction and DMI. (e) and (f) represent the dominant exchange frustration limit, showcasing the CoSS and FM phases across the entire magnetic field range. In this limit, the CoSS phase gradually saturates into the FM phase as the magnetic field increases.}
    \label{Model_Fig}
\end{figure}
To support the findings of Sk-ASkL in the magnet/semiconductor systems, we provide a minimal spin lattice model.
%
%
%
The stability of the Sk-ASkL phase, as determined from our $C_{1v}$ example systems, appears to be crucially dependent on two factors: exchange frustration and the existence of an anisotropic interaction parameter space.
This complex interplay can be effectively represented by a minimal two-dimensional (2D) atomistic model. 
The model incorporates two exchange couplings{--}nearest-neighbor ($J_1$) and next-after-nearest-neighbor ($J_3$) in a square lattice{--}and a nearest-neighbor interfacial DMI ($\textbf{D}$) oriented perpendicular to the bond. 
The square lattice has a lattice constant $a$.
To simplify the model and specifically focus on frustration-driven spin spirals along the (110) direction in a square lattice, we have set $J_2=0$ in our model. This simplification allows us, without loss of generality, to analyze exchange frustration along a specific bond direction.
Without DMI, this model precisely confirms an exchange frustration-driven spin-spiral solution, as previously predicted by Rybakov et al.~\cite{hopfion} within a cubic lattice framework.
The corresponding Hamiltonian, under an external magnetic field $\textbf{B}_\textrm{ext}$, takes the form:
\begin{equation}
	\mathcal{H}=-J_{1}\mathbf{\hat{n}_0}\cdot \mathbf{\hat{n}_1}-J_{3}\mathbf{\hat{n}_0}\cdot \mathbf{\hat{n}_3} + \textbf{D} \cdot (\mathbf{\hat{n}_0}\times\mathbf{\hat{n}_1)}-\mu_s\sum_{i} \textbf{B}_\textrm{ext} \cdot {{\mathbf{\hat{n}}_i}}
	\label{spin_hamiltonian}
\end{equation}
Here, $J_1$ ($>0$) and $J_3$ ($<0$) represent the ferromagnetic and antiferromagnetic exchange constants, respectively. 
The values of $J_3$ and the DMI magnitude $D$ are expressed in units of $J_1$. 
$\mathbf{n}$ is the unit vector along the magnetic moment direction.
We begin by determining the specific conditions under which exchange frustration triggers a spin-spiral state. 
Notably, when exchange frustration occurs along certain bond directions, it drives a spin-spiral solution that propagates along the (110) direction. 
This happens when conditions like $J_3 < -J_1/4$ is satisfied, a phenomenon consistent with the behavior of frustrated spin systems in one dimension. 
The ratio $J_3/J_1$ systematically determines the spin-spiral period; for example, the period decreases from $7.78a$ to $5a$ upon reducing $J_3$ from $-0.3$ meV to $-0.4$ meV.
Moreover, the interplay of $J_3$, $J_1$ and \textbf{D} leads to a cycloidal spin-spiral (CySS) solution, a characteristic feature of isotropic chiral magnets.

We now turn to the second crucial factor: anisotropy in the interaction parameter space. 
To quantify this, we introduce two dimensionless parameters: 
$\gamma_1 = \frac{J_{y1}}{J_{x1}} = \frac{J_{y3}}{J_{x3}}, \quad \text{and} \quad \gamma_2 = \frac{|D_{y1}|}{|D_{x1}|}$,
where \( J_{x1} \) and \( J_{y1} \) denote the first neighbor exchange interactions along the \( x \)- and \( y \)-directions, respectively, while \( J_{x3} \) and \( J_{y3} \) represent the corresponding third neighbor couplings. 
Likewise, \( D_{x1} \) and \( D_{y1} \) refer to the magnitude of DMI along the \( x \)- and \( y \)-directions for first neighbors (See schematic in Fig.~\ref{Model_Fig}).
When $\gamma_1$ and $\gamma_2$ are tuned below unity, it consistently leads to a weaker exchange and DMI in the $y$-direction compared to the $x$-direction.
In physical systems, such anisotropic parameters may arise from specific crystal geometries, like a transition from a square to a rectangular unit cell, or from the morphology of the substrate, as seen in $C_{1v}$ systems.

Figure~\ref{Model_Fig} illustrates how this minimal model, incorporating anisotropic exchange and DMI parameters, accurately depicts the phase transition from the CySS to the Sk-ASkL phase. 
Its broad applicability is further demonstrated by its success in two familiar isotropic limiting cases: the strong DMI limit and the strong exchange frustration limit.
To illustrate the different phases and their transitions as a function of $\textbf{B}_\textrm{ext}$, we selected three specific parameter sets. In all cases, we have set $J_1=1$. 

\subsection{Chiral magnets with strong DMI}
For chiral magnets with strong DMI{--}meaning the DMI is comparable in magnitude to the exchange parameters{--}, we have fixed $J_2 =-0.3$ and $|\textbf{D}|=0.2$. 
The resulting energy landscape of stable spin textures is shown in Fig.~\ref{Model_Fig}(a).

In the absence of an external magnetic field, the system's ground state is typically a CySS.
With the introduction of an external magnetic field ($B_\textrm{ext}$), a first-order phase transition enables the stabilization of a SkL as the ground state. 
This robust SkL phase is sequentially observed between the initial CySS state and a subsequent ferromagnetic (FM) state.
This phenomenon is commonly observed in interfacial chiral magnets, such as heterostructures composed of transition metals and heavy metals (e.g., Fe/Ir(111), Mn/W(001) systems).
Notably, neither the Sk-ASkL nor the CoSS can be stabilized in this regime due to the strong DMI, which energetically disfavors their formation.

We further investigate the effect of anisotropic DMI, specifically when $D_{x1}\neq D_{y1}$, by modifying the parameter $\gamma_2$.
Our findings reveal a critical dependence of the SkL phase on this DMI anisotropy parameter.
Below a certain critical value of $\gamma_2$, the SkL phase loses its energetic favorability, leaving the CySS and FM phases as the only stable solutions.
This is analogous to the behavior observed in monoaxial chiral magnets, characterized by the absence of DMI along one direction, as described recently in Ref.~\cite{monoaxial}.

\subsection{Anisotropic chiral magnets with the emergence of the Sk-ASkL}
Now, we investigate the model for a significantly reduced DMI of $|\textbf{D}|=0.02$ (an order of magnitude smaller) while holding $J_3$ constant at $-0.3$.
It is important to note that exchange frustration exerts a greater influence than the DMI in this setup.
For the isotropic limit ($\gamma_1 = \gamma_2 =1$), we observe two phase transitions: a CySS state that transitions into a conical spin-spiral (CoSS) state, which then transitions into an FM state. 
As depicted in Fig.~\ref{Model_Fig}(c), while both SkL and Sk-ASkL states can be formed, they are metastable configurations, residing at higher energy levels.

We now consider anisotropic exchange interaction and DMI by setting $\gamma_1=\gamma_2=0.1$, effectively weakening interactions along the $y$-axis.
This anisotropy plays a significant role in tuning the relative energies of the competing phases.
In particular, DMI anisotropy increases the energy of both the SkL and CySS phases, while having no effect on the Sk-ASkL energy. 
On the other hand, exchange anisotropy influences all phases but impacts the CoSS phase more strongly than the Sk-ASkL. 
As a result, the combined effect of exchange and DMI anisotropy leads to the stabilization of Sk-ASkL as the ground state within a considerable field range.
Figure~\ref{Model_Fig}(d) presents this magnetic phase evolution with increasing external field, showing a clear sequence of transitions: CySS$\rightarrow$Sk-ASkL$\rightarrow$CoSS$\rightarrow$FM.

\subsection{Strong exchange frustration limit}
To explore the regime where exchange frustration outweighs the DMI, we set the parameters \( J_3 = -0.4 \) and $|\textbf{D}|=0.002$. 
This parameter set significantly enhances the relative influence of frustrated exchange interactions compared to DMI within our model.
Corresponding results for the isotropic and anisotropic scenarios are depicted in Figs.~\ref{Model_Fig}(e) and (f), respectively.
Remarkably, the CoSS and FM phases are identified as the lowest energy equilibrium states across the entire range of non-zero magnetic fields.
The dominant exchange frustration mechanism preferentially stabilizes the CoSS phase below a critical field. 
In this regime, all other non-collinear phases exist as metastable states. 
As the magnetic field increases beyond this threshold, the CoSS phase smoothly evolves into the FM phase.
This behavior, dominated by exchange frustration, is consistent with previous findings reported in Ref.~\cite{Leonov2015}.
To ensure completeness, we validated our model by considering anisotropic limit, $\gamma_1=\gamma_2=0.1$.

Within the minimal $J_1$-$J_3$-$D$ model, we conclude that the formation of the Sk-ASkL as a ground state critically depends on two essential conditions: (i) a balance between exchange frustration and DMI energies, and (ii) the existence of anisotropy in both the exchange interaction and DMI. 
Notably, this framework can be extended to systems with long-range frustrated exchange interactions and DMI, which is characteristic of the real materials discussed in our study.

\subsection{Sk-ASkL stability within anisotropic exchange}

The \textit{net-zero} lattice configuration, containing an equal number of skyrmions and antiskyrmions, becomes a ground state due to the anisotropy in DMI and exchange interactions.
To elucidate this, we qualitatively explain how these two key features govern the energy balance among the noncollinear phases: CySS, CoSS, Sk-ASkL, and SkL.
%
%
%
Notably, the anisotropic DMI plays a critical role in mediating the energy balance specifically between the SkL and Sk-ASkL phases. This is stemming from the distinct magnetization properties of these topological configurations.
For instance, the homogeneous chirality of both CySS and SkL means that any reduction in DMI due to its anisotropic nature along a crystallographic direction directly increases their energy (see Fig.~\ref{ani_DMI}).
%
%
%
On the contrary, the magnetic texture in Sk-ASkL does not show homogeneous chirality due to the presence of antiskyrmions.
%
A distinctive property of the Sk-ASkL is its chirality, which is consistently maintained along the $x$-axis but alternates along the orthogonal ($y$-axis) direction.
This alternating chirality leads to an average vanishing of the DMI energy contribution. 
Consequently, the DMI's influence on the S-AL phase energy becomes largely independent of the DMI strength along the $y$-axis.
%
%
This is also true in the case of CoSS, where varying the DMI coupling strengths in any of the two orthogonal directions does not affect the DMI energy, as this phase is achiral in nature.
Thereby, with anisotropy in DMI, we increase the energy of CySS and SkL while the energy of Sk-ASkL and CoSS remains almost unchanged as presented in Fig.~\ref{ani_DMI}.

Next, the anisotropy in exchange frustration plays a crucial role in tuning the energy balance between the Sk-ASkL and CoSS.
Since the CoSS phase originates solely from exchange frustration, any anisotropy effectively reduces its exchange energy contribution, thereby increasing its energy more significantly than that of Sk-ASkL.
In this way, we can switch the energy balances and make Sk-ASkL energetically more favorable stable phase than the other phases.
\begin{figure}
    \centering
    \includegraphics[width=0.6\linewidth]{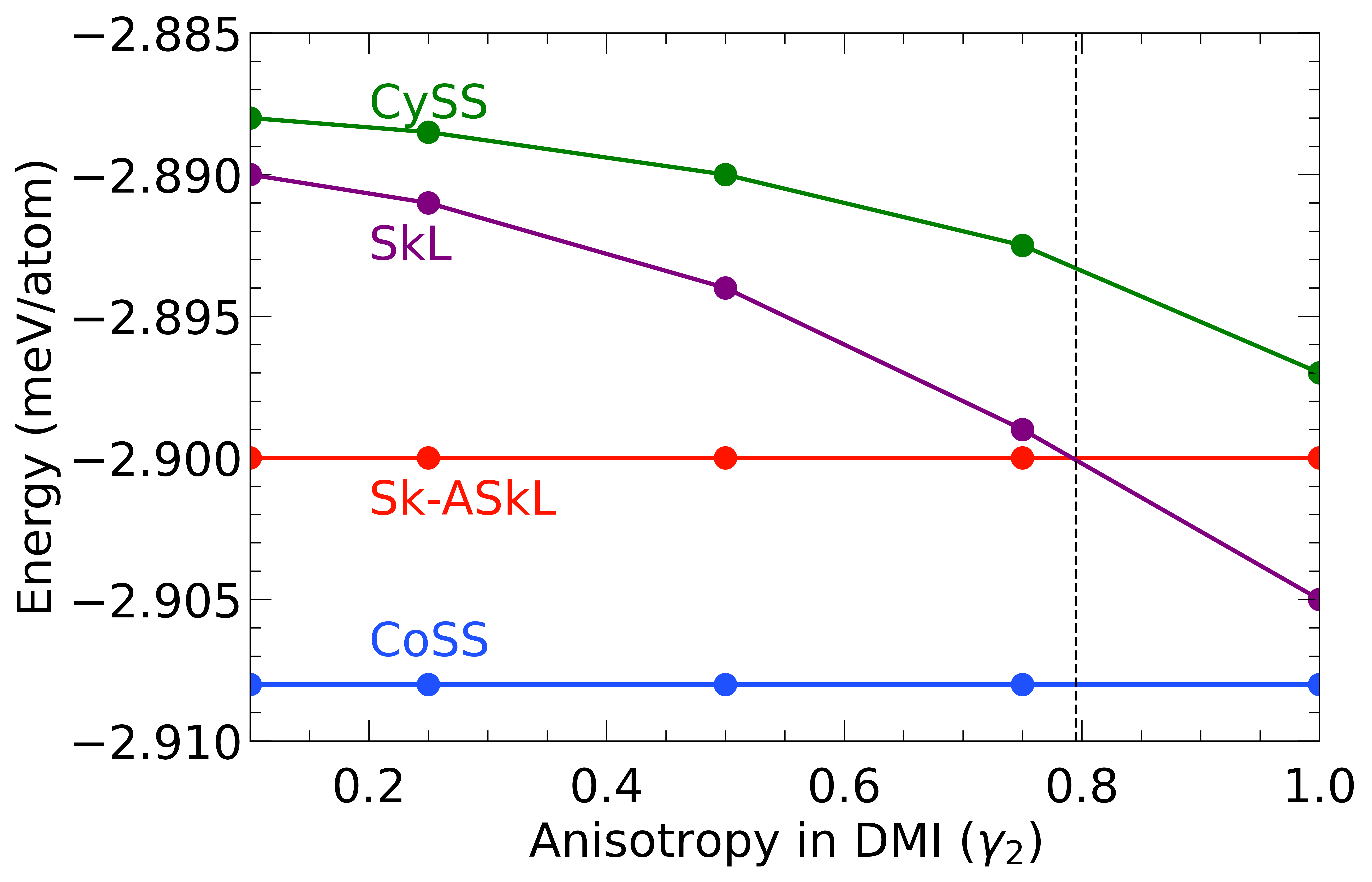}
    \caption{Effect of anisotropy in DMI on different magnetic phases while keeping isotropic exchange interaction. The zero magnetic field state, i.e., CySS, is found to increase its energy due to the reduction of DMI along the $ y$-direction. At a finite magnetic field of about 55 mT, energies of SkL, Sk-ASkL, and CoSS phases are shown. Both Sk-ASkL and CoSS phases remain unchanged in energy with anisotropy in DMI. Below a critical value of $\gamma_2\approx0.8$, the Sk-ASkL solution becomes lower in energy than the SkL.}
    \label{ani_DMI}
\end{figure}
\section{Results for 2F\lowercase{e}/C\lowercase{d}T\lowercase{e}(110) system from KKR calculations}

A class of systems exhibiting this $C_{1v}$ symmetry includes magnetic films (transition metals) on the (110) surface of zincblende semiconductors.
In this context, 2Fe/CdTe(110) is another prototypical experimental system we consider. 
The selection of CdTe, a high atomic number semiconductor substrate, is motivated by the potential for a strong DMI. 
We have systematically studied this system using the same approach as for 2Fe/GaAs(110).

\begin{figure}{!bh}
    \centering \includegraphics[width=0.75\linewidth]{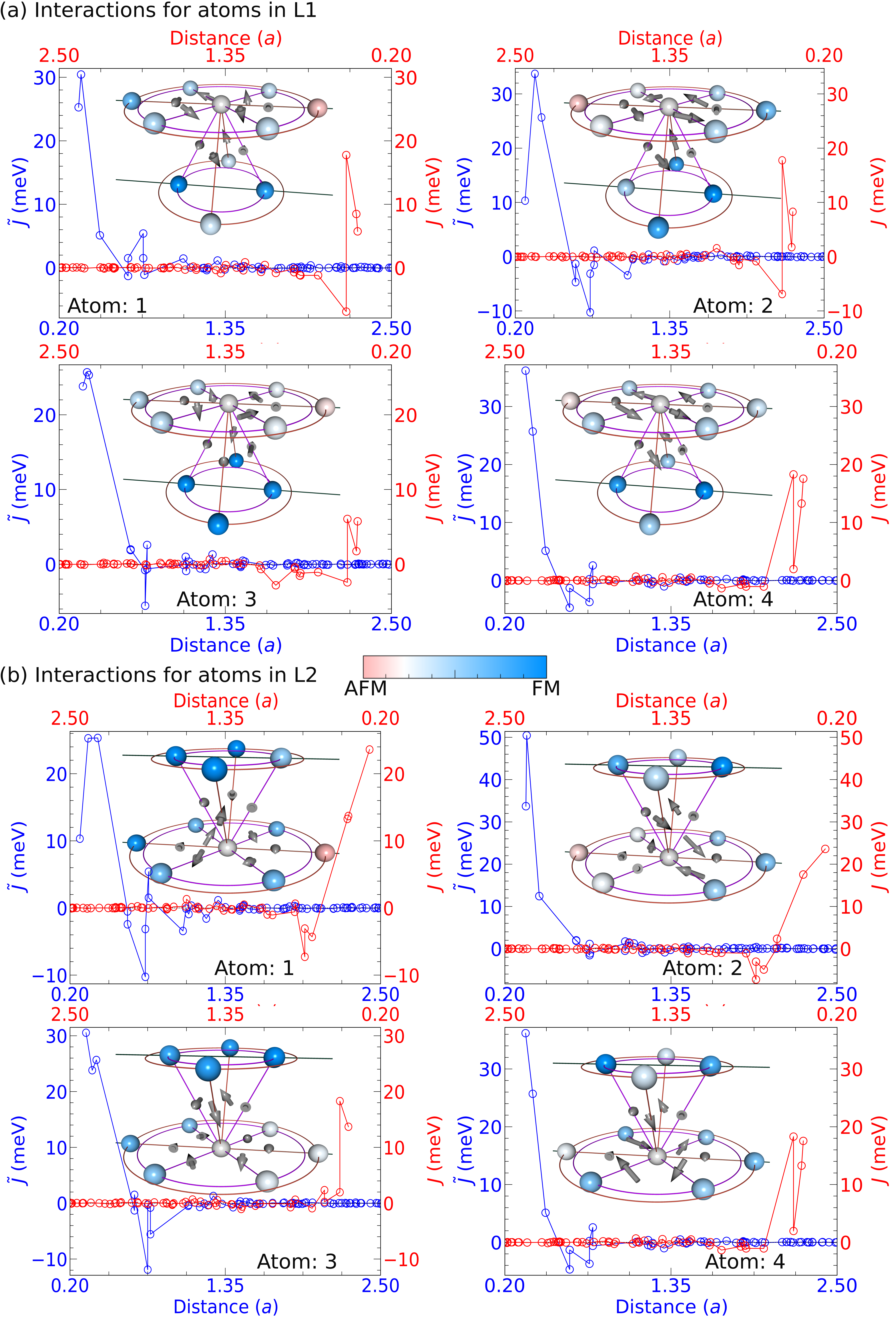}
    \caption{Magnetic interactions in the 2Fe/CdTe(110) system are illustrated, showing interactions for atoms in layer L1 in (a) and layer L2 in (b).Interlayer and intralayer exchange interactions are represented by blue and red lines, respectively. Inset arrows indicate the orientation of DM vectors for each atom. Also, the magnitude and type of exchange energy are represented by colored spheres, where color saturation indicates strength, and blue/red denote ferromagnetic/antiferromagnetic interactions. }
    \label{CdTe_Int}
\end{figure}

\subsection{Anisotropic magnetic interaction parameters}
The magnetic moments of Fe atoms on the CdTe(110) surface are shown in Table \ref{MagMom_CdTe}.
As expected, these moments vary depending on the atomic site. However, for the spin-lattice simulations, an average Fe moment of approximately 2.70 $\mu_\textrm{B}$ is used. 
Notably, the MCA here is also out-of-plane and stronger than that observed for 2Fe/GaAs(110).

The exchange interactions and DM vector orientations are shown in Fig. \ref{CdTe_Int}. 
The maximum intralayer and interlayer exchange strengths are 23.7 meV and 50.4 meV, respectively. 
This strong interlayer exchange is attributed to the reduced interlayer distance between Fe layers, a consequence of the significant tensile strain (approximately 11.4\%) resulting from the large lattice mismatch. However, compared to 2Fe/GaAs(110), the exchange frustration is considerably stronger, and we again observe a frustration-driven SS state with a period of $\lambda \sim$ 2.6 nm. 
The DMI exhibits distinct rotational senses: left-rotating in layer L1, right-rotating in layer L2, and left-rotating interlayer. 
The resulting DMI rotation is left-rotating, consequently leading to a right-rotating CySS state with a period of $\lambda \sim$ 2.3 nm as the zero-filed ground state.

\begin{table}[h]
	\centering
	\begin{tabular}{|c|c|c|c|c|}\hline
	Magnetic layers&Fe atom& Magnetic mom. in $\mu_B$&Average ($\mu_B$)&K (meV/Fe atom) \\ \hline
    \multirow{4}{*}{Fe layer 1}
        &Fe 1&2.78&& \\ \cline{2-3}
        &Fe 2&2.84&&  \\ \cline{2-3}
        &Fe 3&2.92&&  \\ \cline{2-3}
        &Fe 4&2.92&2.70&0.7  \\ \cline{1-3}
    \multirow{4}{*}{Fe layer 2}
        &Fe 1&2.49&&  \\ \cline{2-3}
        &Fe 2&2.36&&  \\ \cline{2-3}
        &Fe 3&2.74&&  \\ \cline{2-3}
        &Fe 4&2.59&&  \\ \hline
        \end{tabular}
	\caption{Magnetic moments of Fe atoms in the magnetic layer and the MCA parameter.}
	\label{MagMom_CdTe}
\end{table}

\subsection{Spontaneous generation of $Q=\pm 1$ solitons and the equilibrium phases}
This section details the thermodynamic stability of metastable states characterized by the coexistence of Sks and ASks.
The magnetization evolution of 2Fe/CdTe(110) under an applied external magnetic field ($B_{\mathrm{ext}}$) was tracked using controlled thermal protocols. 
Specifically, the finite-field spontaneous nucleation process involved cooling the magnetic domain at a fixed $B_{\perp}$ = 1 T. 
The initial temperature was 30 K, with a random spin configuration (see the Supplementary Movie).
Figure~\ref{sponte} provides snapshots corresponding to four temperatures: 30 K, 20 K, 10 K, and 0 K.
The system was initially relaxed at the starting temperature for $10^4$ MC steps.
Then, at each successive temperature step of $\Delta T$ = 2.5 K, we have applied $10^5$ relaxation steps.
As shown in the Supplementary Movie, two distinct spin configurations spontaneously nucleate. 
These configurations exhibit topological charges of $-1$ (Sk) and $+1$ (ASk). 
We performed several temperature cycles involving consistent heating and cooling.
We consistently observe elongated Sks and ASks, as indicated by the red and magenta boxes in the T = 0 K snapshot in Fig.~\ref{sponte}(d). 
Remarkably, this unique coexistence of topologically opposite quasiparticles, along with the conical background modulation, is stable without annihilation.

Despite the stability of individual Sks and ASks in these frustrated chiral magnets, the competing Sk-ASkL and SkL phases are energetically close. 
Notably, here the antiskyrmion lattice consistently exhibits a higher energy state, which we attribute to the interfacial $C_{nv}$-type DMI textures.
To simulate these lattice configurations, we employ energy minimization, initializing the simulation domain with regular hexagonal arrangements corresponding to SkL and Sk-ASkL configurations. 
Starting from 50 K, we performed simulated annealing with steps $\Delta T = 2$~K for equilibration with OBC.
After the annealing process, the hexagonal lattices are found to minimize their energy in an elongated hexagonal form.  
In this setup, the simulation domain contains the same number of spin textures for both SkL (comprising only skyrmions) and Sk-ASkL (comprising equal numbers of skyrmions and antiskyrmions). 
Crucially, the Sk-ASkL demonstrates a lower energy state than the SkL across a broad range of magnetic field values. 
Furthermore, the robustness of the Sk-ASkL phase was observed under multiple thermal cycles (heating and cooling within 20 K windows) with both PBC and open OBC. 
It remained the most energetically favorable state within a certain field range.

\begin{figure}[h]
    \centering
    \includegraphics[width=0.9\linewidth]{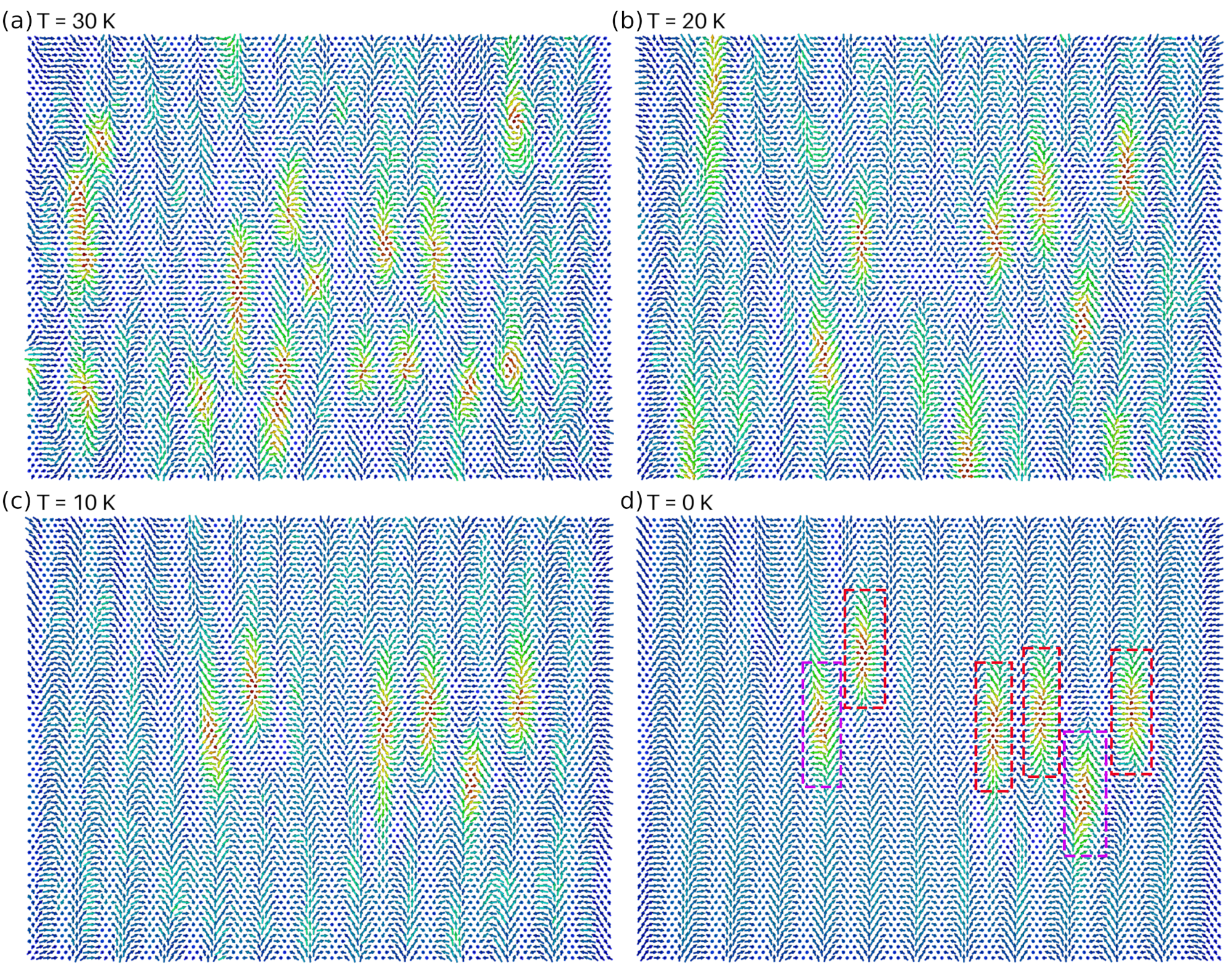}
    \caption{Spontaneous nucleation of skyrmions and antiskyrmions in the presence of a perpendicular magnetic field, $B_\perp$ = 1 T. Snapshots of the system are shown for temperatures (a) 30 K, (b) 20K, (c) 10K, and (d) 0 K. The Sks and ASks are indicated by red and magenta boxes, respectively.}
    \label{sponte}
\end{figure}

Figure~\ref{CdTe} shows the spin configurations of all equilibrium phases calculated for the 2Fe/CdTe(110) system. 
A unique feature of this class of magnets is the CoSS phase, which appears between the Sk-ASkL phase and the saturated ferromagnetic phase.  

\begin{figure}[H]
    \centering
    \includegraphics[width=1.0\linewidth]{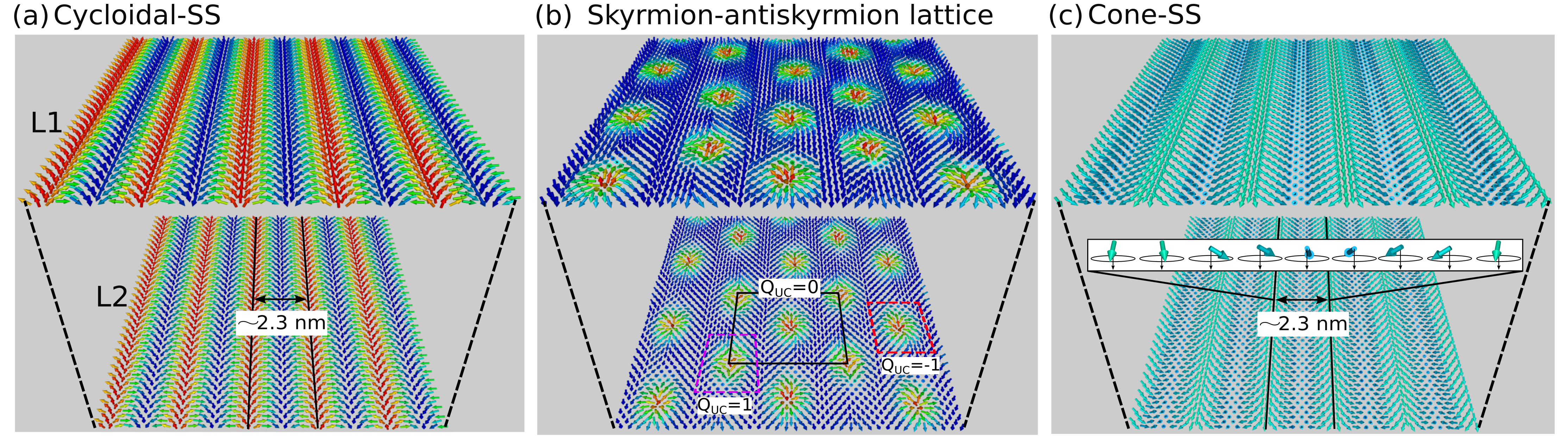}
    \caption{Equilibrium spin textures: (a) right-rotating CySS at zero field, (b) the Sk-ASkL and (c) CoSS phases at finite field. The antiskyrmion (skyrmion) is highlighted with a magenta (red) box. The unit cell topological charge, $Q_\textrm{UC}$=0, defines a \textit{net-zero} topological lattice. The inset depicts an enlarged angular view of the selected CoSS spins.}
    \label{CdTe}
\end{figure}
\section{Dynamical behavior: Current Driven motion}
To demonstrate the advantage of this novel phase, we have included the current-induced dynamics of an isolated pair. 
It is important to clarify that our primary results are established in Density Functional Theory and atomistic spin dynamics. 
The requisite conditions for stabilizing this phase have been meticulously identified through investigations of real materials and a generic $J_1$-$J_3$-$D$ model. 
Subsequently, our micromagnetic simulations, which incorporated these critical conditions, have unequivocally validated Sk-ASkL as the lowest energy solution within a finite magnetic field window. 
Further micromagnetic simulations, employing the continuum model within MuMax~\cite{mumax}, provide a robust framework for investigating the current-induced dynamics of magnetic textures.
Our model incorporates exchange frustration through the inclusion of second- and fourth-order exchange terms, in conjunction with DMI and an external magnetic field. 
The energy functional is given by:
\begin{align}
    E(\mathbf{n})= \int\left(\mathcal{A}\Biggl[\,\left(\dfrac{\partial \mathbf{n}}{\partial x}\right)^{2} \,+\,\, \left(\dfrac{\partial \mathbf{n}}{\partial y}\right)^{2}\,\Biggr] 
 + \mathcal{B}\Biggl[\left(\dfrac{\partial^2 \mathbf{n}}{\partial x^2}\right)^{2} + \left(\dfrac{\partial^2 \mathbf{n}}{\partial y^2}\right)^{2}\Biggr] 
 +\mathcal{D}\Biggl[\Lambda^{(x)}_{xz}+\Lambda^{(y)}_{yz}\Biggr]-\mathcal{K}n^2_z-M_s\mathbf{B}_\mathrm{ext}\cdot\mathbf{n} \right)d\mathbf{r}
    \label{A_24_D1}
\end{align}
where $\mathcal{A}$ and $\mathcal{B}$ are the coefficients of second- and fourth-order terms, respectively; $\mathcal{D}$ is the magnitude of DMI; $\mathcal{K} (>0)$ is out-of-plane anisotropy. 
The last term is the Zeeman term with an external magnetic field $B_\mathrm{ext}$, $n=\mathbf{M}/M_s$ with $M_s$ being the saturation magnetization.
The Lifshitz invariants are defined as $\Lambda_{ij}^{(k)} = n_i \frac{\partial n_j}{\partial r_k} - n_j \frac{\partial n_i}{\partial r_k}$ which defines the chirality of the DMI.
An exchange-frustration-driven spin spiral can be achieved at zero magnetic field, even without DMI, by considering $\mathcal{A}<0$ and $\mathcal{B}>0$. 
This observation aligns with findings reported in Ref.~\cite{hopfion}.

We consider a rectangular domain (racetrack) with $L_x=300$ nm and $L_y=100$ nm with a mesh density of $384\times128\times1$. 
In line with standard practice, we define dimensionless magnetic field $\mathbf{h}=\mathbf{B_\textrm{ext}}/B_c$, anisotropy $u=\mathcal{K}/{M_sB_c}$.
The parameter $B_c=\mathcal{A}^2/4\mathcal{B}M_s$ is the critical field for the saturation magnetization.
The simulations are performed with the following parameters: $\mathcal{A}=-10^{-17}$ J/m, $\mathcal{B}= 10^{-34}$ J.m, $\mathcal{D}=-10^{-10}$ J/m$^2$, $h=2.5$, $u=1.5$. 
The parameters for the damping and nonadiabatic torque are set to  $\alpha=0.001$ and $\xi=0.01$, respectively. 

\begin{figure}
    \centering    \includegraphics[width=1.0\linewidth]{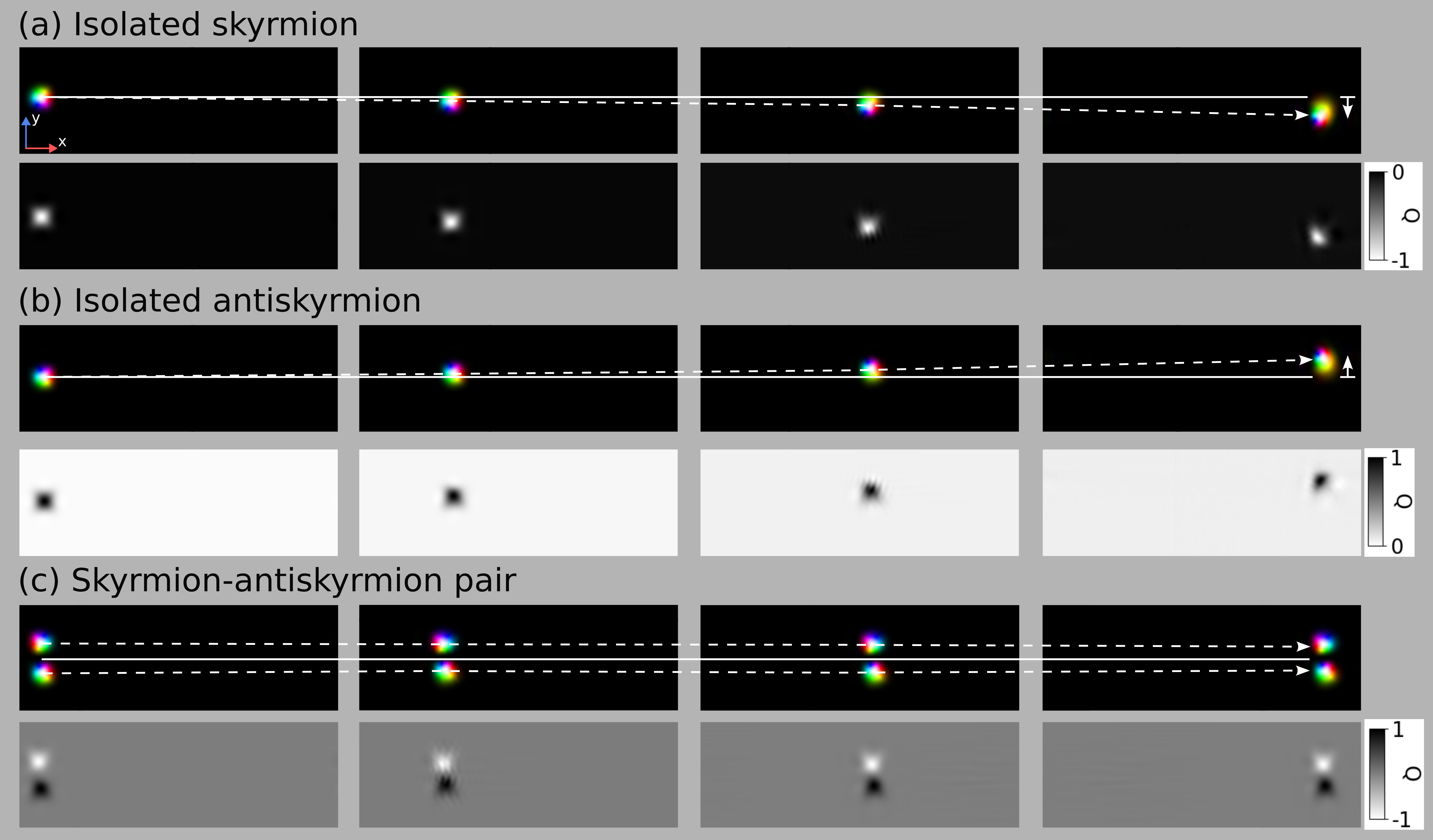}
    \caption{Current-driven dynamics are depicted through snapshots showing the motion of (a) an isolated skyrmion, (b) an isolated antiskyrmion, and (c) a skyrmion-antiskyrmion pair.}
    \label{hall}
\end{figure}

First, we stabilize the spin configurations, i.e., an isolated skyrmion, an isolated antiskyrmion, and a skyrmion-antiskyrmion pair within the racetrack through energy minimization.
This minimization guarantees the uniformity in size and shape across all magnetic textures.
Following this, an in-plane current, set at a density of $j=3\times10^9$ A/m$^2$, is applied along the $x$ direction.
Snapshots taken at different time intervals, as shown in Fig.~\ref{hall}, depict the resulting current-driven motion of these three distinct spin configurations.
Consistent with prior research and the established understanding of skyrmion current dynamics, isolated skyrmions driven by an in-plane current exhibit a skyrmion Hall effect, deflecting transversely along the $-y$ direction (see Fig.~\ref{hall}(a)).
As expected due to their inverse topological charge, antiskyrmions exhibit the opposite behavior, deflecting transversely along the $+y$ direction (Fig.~\ref{hall}(b)).
Remarkably, as shown in Fig.~\ref{hall}(c), the skyrmion-antiskyrmion pair exhibits no transverse deflection under current.
This phenomenon occurs because the Magnus forces acting on the skyrmion and antiskyrmion are precisely equal and opposite, resulting in their cancellation. 
Consequently, the skyrmion-antiskyrmion pair moves without experiencing the Skyrmion Hall effect.

The well-established approaches for suppressing the Skyrmion Hall Effect involve utilizing synthetic antiferromagnetic bilayers.
In these systems, a net-zero topological charge is achieved by combining two skyrmions residing on antiferromagnetically coupled ferromagnetic layers. 
Due to their opposite topological charges, these coupled skyrmions experience mutually opposing Magnus forces, which cancel each other. 
This enables the composite net-zero skyrmion to move precisely along the applied current direction. 
As highlighted, relevant work by Ezawa et al.~\cite{Zhang_NatCommun2016} and Du et al.~\cite{Zhang_NanoLett2024}, which indeed showcase Skyrmion Hall Effect-free spin textures achieved through geometrical bilayer constructions. 
While conventional methods for achieving Hall-free motion typically require multi-layered structures and precise interlayer coupling, our Sk-ASkL system uniquely offers this functionality within a single magnetic layer. 
This represents a novel alternative approach, where objects composed of topologically opposite quasiparticles form a net-zero bag, thereby inherently minimizing transverse deflection and enabling Hall-effect-free motion.

\end{onecolumngrid}

\end{document}